\documentclass{aa}  
\usepackage{natbib}
\usepackage{subcaption}
\usepackage{graphicx}
\usepackage{txfonts}
\usepackage[switch]{lineno}
\usepackage{xcolor}

\newcommand{\g}{\ensuremath{\gamma}\xspace}
\newcommand{\hd}{H$_2$\xspace}
\newcommand{\hi}{\ion{H}{i}\xspace}
\newcommand{\alfven}{Alfv\'{e}n\xspace}
\newcommand{\alfvenic}{Alfv\'{e}nic\xspace}

\newcommand{\cmsqs}{cm$^2$ s$^{-1}$\xspace}

\begin{document}

   \title{Cosmic-ray diffusion and the multi-phase interstellar medium in a dwarf galaxy}

   \subtitle{I. Large-scale properties and \g-ray luminosities}

   \author{A. N\'u\~nez-Casti\~neyra
          \inst{1}, I. A. Grenier\inst{1}, F. Bournaud\inst{2}, Y. Dubois\inst{3},
          F. R. Kamal Youssef\inst{1}
          \and
          P. Hennebelle\inst{2}
          }

   \institute{Universit\'{e} Paris Cit\'{e} and Universit\'{e} Paris Saclay, CEA, CNRS, AIM, F-91190 Gif-sur-Yvette, France
        \and
            Universit\'{e} Paris Saclay and Universit\'{e} Paris Cit\'{e}, CEA, CNRS, AIM, F-91190 Gif-sur-Yvette, France
        \and 
             Institut d’Astrophysique de Paris, UMR 7095, CNRS and Sorbonne Universit\'{e}, 98 bis boulevard Arago, 75014 Paris, France
             }

   \date{Submitted May 16,2022; accepted xx xx,xxxx}

 
  \abstract
  {Dynamically, cosmic rays with energies above about one GeV/nucleon may be important agents of galaxy evolution. 
  Their pressure and pressure gradients compare with the thermal and magnetic ones to alter gas accretion onto a galaxy, drive fountains and massive galactic outflows, and alter the mass cycling between the diffuse gas reserves and the dense clouds where stars form. The feedback efficiency depends on the actual properties of cosmic-ray transport in the different media, so we crucially need theoretical clues and observational constraints on these properties in order to assess the impact of cosmic rays on galaxy evolution.}
  {We aim to study the dynamical role of cosmic rays in shaping the interstellar medium of a galaxy when changing their propagation mode. As a first step, we perform high-resolution simulations of the evolution of the same isolated galaxy and compare the impact of the simplest cosmic-ray transport assumption of uniform diffusion. We also compare the total \g-ray luminosity produced in the galaxy by hadronic interactions between cosmic rays and the gas as this observable encapsulates the convolution of the gas and cosmic-ray spatial distributions and it can be compared to Fermi LAT data.}
  {We have simulated the evolution of a gas-rich dwarf galaxy (${\sim}10^{11}$ M$_{\odot}$ in total mass, forming about 1 M$_{\odot}$~yr$^{-1}$ of stars) with the magnetohydrodynamic adaptive-mesh-refinement code RAMSES, down to 9-pc resolution. Cosmic rays are advected by the gas and they diffuse either isotropically or preferentially along the magnetic field with uniform diffusion coefficients ranging from $3\times 10^{27}$ to $10^{29}$ cm$^2$ s$^{-1}$ in order to bracket the average value inferred in the Milky Way. The simulations use gas cooling and heating functions that model the multiphasic structure of the gas down to a much lower resolution than used here. We have also updated the observational relation seen between the \g-ray luminosities and star-formation rates of galaxies using the latest detection and characterisations of Fermi LAT sources.}
  {We focus this article on the impact of CR transport on the large-scale properties of the galaxy. We find that the radial and vertical distributions of the gas in the different phases, as well as the mass ranking between the phases, are marginally altered when changing cosmic-ray transport. We observe a positive feedback of cosmic rays on the amplification of the magnetic field in the inner half of the galaxy, except for fast isotropic diffusion. The increase in cosmic-ray pressure for slow or anisotropic diffusion can suppress star formation by up to 50\%, but the dual effect of cosmic-ray pressure and magnetic amplification can reduce star formation by a factor 2.5.
  The global \g-ray luminosities and star-formation rates of the simulated galaxies are fully consistent with the best-fit trend seen in the observations in the case of anisotropic $10^{27.5-29}$ cm$^2$ s$^{-1}$ diffusion and for isotropic diffusion slower or equal to $3 \times 10^{28}$cm$^2$ s$^{-1}$. These results therefore do not confirm claims of very fast $10^{29-31}$ cm$^2$ s$^{-1}$ diffusion to match the Fermi LAT observations.} 
   {}

   \titlerunning{Cosmic-ray diffusion in the multi-phase ISM I}
   \authorrunning{N\'u\~nez-Casti\~neyra et. al.}
   \maketitle
\section{Introduction}
The stellar content of a galaxy is the result of the competition between the gravitational collapse of cold-dense gas, magnetic resistance, and a series of energetic feedback processes. For low-mass galaxies, the predominant process is supernova (SN) feedback  \citep{Dekel1986}. Numerical simulations of such galaxies have thus been focusing on achieving effective numerical description of SN feedback to reproduce galactic observations \citep[e.g.][]{Stinson2006,Teyssier2013,Springel2003,Dubois2008,DallaVecchia2012,Kimm2015}. 
Other feedback processes like stellar radiation \citep{Hopkins2011,Agertz2013,Aumer2013,Rosdahl2015,Emerick2018}, pressure from Lyman-$\alpha$ photons \citep{Kimm2018}, and cosmic rays (CRs) are now suspected to be important processes to modulate star formation. 
Simulations showed that CRs can thicken galactic discs and provide significant pressure support to drive galactic winds out of dwarf galaxies \citep{Booth2013,Ruszkowski17,Dashyan20} and out of Milky-Way size galaxies \citep{Salem2014,Farcy2022}. Changes in CR transport can affect the mass loading and the gas state in the outflows \citep{Girichidis18}.

Low-energy CRs, with kinetic energies around one GeV per nucleon, contain most of the energy and momentum of the particle population. At least in the Milky Way, the CR energy densities compare with those of the turbulent and magnetic components \citep{Wefel1987, Zweibel2017}, so CR pressure gradients can influence the gas dynamics, cloud formation and galactic-wind outflows. The dynamical role of CRs was first recognized by Parker (1966) who noted that, in a vertically stratified ISM, the magnetic and CR pressures would inflate the thermal gas and excite a vertical oscillation of the gas layer, named the Parker instability. Recent studies show that the instability growth rate depends on CR transport \citep[see section 6.4 of][]{Hanasz21}.

Low-energy CRs behave as a relativistic fluid with a soft equation of state. They are thus less prone to pressure losses due to adiabatic expansion and they increase the gas compressibility. CR nuclei have long cooling times of $10^{6-7}$~Myr mostly due to pion production in hadronic interactions with the interstellar gas \citep{Schlickeiser09}, so their energy density lingers in the ISM for longer times than the thermal one. CRs diffuse away from their production site with diffusion lengths exceeding the typical size of star-forming clouds \citep{Acero16diff}, so they should not dynamically affect individual clouds on scales of several tens of parsecs, but their large-scale pressure gradients can modify the gas circulation and condensation from the broader diffuse gas reservoirs to dense cloud complexes. 

The main CR component, $\sim$GeV protons and alpha particles, exchange energy collisionlessly with the thermal gas, through the local magnetic field, kinetic scale waves, and gyro-resonant streaming instabilities \citep{Kulsrud2005}. 
CRs scatter off magnetic fluctuations that they can excite via resonant streaming instability ("self-confinement") or that are induced by interstellar \alfvenic turbulence ("extrinsic confinement") driven by stellar activity and cascading down to the micro-parsec scale of CR gyroradii (see reviews by \cite{Zweibel2017} and \cite{Hanasz21}). In the Milky Way, breaks in the observed CR spectra suggest a possible transition between self-confinement in the disc and extrinsic diffusion at larger heights (few kpc) above the disc \citep{Blasi2012,Evoli2018}, but \g-ray observations of the vertical CR-flux gradient do not support this scenario \citep{Joubaud20}. 

At GeV energies, the dominant transport mode is likely to be self-confinement via the streaming instability \citep{Kulsrud69,Plotnikov21}, but damping processes can inhibit the growth of the instabilities. Depending on the local damping rate, CRs can be efficiently scattered by strong resonant \alfven waves, thus be well isotropised and stream at a velocity close to the local ion \alfven speed. Conversely, they can be loosely coupled to the waves, scatter off them, and diffuse at velocities far exceeding the ion \alfven speed \citep{Kulsrud71}. The CR distribution is then quite anisotropic and the diffusion coefficient depends on the tangling of the magnetic field in the interstellar MHD turbulence \citep{Xu22}. 
Advection by the flowing gas can also play an important role in CR transport. 

In the low-density, hot ionised gas that fills the disc corona (aka halo), the \alfven wave growth is mostly (but weakly) limited by non-linear Landau damping. CRs primarily are advected away by fast winds and/or by streaming at the large \alfven speeds that prevail in the tenuous gas \citep{Armillotta2022}. 
In the warm ionised medium that extends to 1 or 2 kpc about the plane, turbulent damping resulting from interactions between self-excited \alfven waves and background MHD waves can significantly suppress the streaming instabilities \citep{Farmer2004,Xu22} even though it is often ignored in simulations. The CR diffusion coefficient should then strongly depend on the ambient \alfvenic Mach number \citep{Xu22}. 
Most of the gas mass lies in the denser neutral phases, warm and cold, which extend to a few hundred parsecs about the plane and can be found in higher-altitude filaments entrained in the fountains and winds. In this weakly ionised medium, the \alfven waves are efficiently damped by collisional friction between ions and neutrals. CR stream at large speeds along field lines and the diffusion coefficient should again strongly depend on the ambient \alfvenic Mach number \citep{Xu22}. By post-processing 4-pc to 8-pc resolution simulations of the ISM to solve the CR transport equations, \cite{Armillotta2022} find that the CR mean free path can vary by five orders of magnitude across the different gas states. Further complications arise in the \hi and \hd gas phases at parsec and sub-parsec scales. They are due to the complex filamentary/clumpy structures of the clouds, exhibiting large contrasts in density, ionisation fraction, and magnetic field orientation. The competing changes in ion \alfven speed (downward with increasing density, but upward with decreasing ionisation), in CR pressure gradients (sharp bottlenecks upstream of gas clumps and uniform shadows inside the clumps and behind them), and in the entanglement of the magnetic field, all cast large uncertainties on the effective  transport properties of $\sim$GeV CRs in these environments \citep{Skilling76,Schlickeiser16,Ivlev18,Bustard21}. Another complication arises in the vicinity of CR sources (SN remnants) around which the steep CR density gradient leads to the development of strong streaming instabilities. CR diffusion severely slows down within a few hundred parsecs from the source \citep{Ptuskin08,Semenov21}. 

Conversely, while theory suggests that CR transport parameters should dramatically vary across gas phases and with the ambient magnetization level, the ensemble of CR observations (local CR spectra, secondary-to-primary ratios, \g-ray and radio intensities) can be explained to first order by assuming uniform and isotropic CR diffusion across the entire Milky Way \citep[see reviews by][]{Grenier15,Hanasz21}. The good linear relation that is observed between \g-ray fluxes and total gas column densities in many directions across the Galactic disc implies that variations in CR density are at best modest (a smooth radial gradient by a factor of order four across the Milky Way), and that they are largely decoupled from the gas density and from star-formation activity (e.g. spiral arms, see Fig. 4 of \cite{Grenier15}). 
Furthermore, CR spectra inferred in \g rays hardly soften from the turbulent inner regions to the quieter periphery of the Milky Way \citep{Acero16diff} whereas the energy scaling of the CR diffusion coefficient should depend on the MHD turbulence ratio ($\delta B/B$) \citep{Reichherzer22}. 

Given these theoretical limitations and puzzling data, we need to find observational clues about CR propagation in different media. While galaxy simulations are used to explore the role of CRs on the mass loading of galactic winds, on pressure gradients in the circumgalactic medium, and on star formation across the sequence of star-forming galaxies \citep[e.g.][]{Hopkins2021,Crocker21a,Werhahn21}, we aim to use them to study the degree of coupling between CRs and the ISM at sub-galactic scales and to study the detailed impact of CRs on the different gas phases in Milky-Way-like environments. Characterising the ISM response to CR gradients for different transport assumptions can guide \g-ray observations to constrain CR diffusion properties. 

These goals require CR-inclusive simulations of galaxies with a reliable description of the multiphasic structure of the ISM down to cold and dense clouds on scales of tens of parsecs. The detailed modelling of higher-resolution clouds and of the chemical transition to molecular gas is not required as CR transport through clumpy neutral gas becomes even more complicated, energy-dependent, and uncertain \citep{Bustard21}, and as \g-ray observations can only infer CR fluxes in the \hi-line emitting atomic phase. The uncertainty in the derivation of \hd column densities prevents CR-flux measurements in the molecular gas \citep{Grenier15,Remy17}. 

Most simulations assume some combination of diffusive and streaming CR transport in addition to advection with the thermal gas. But the theoretical understanding of the transition between physics at gyroradii scales and CR transport properties on macroscopic interstellar scales remains poorly understood \citep[e.g.][]{Hopkins21gam,Semenov21}. The bulk diffusion speed and the degree of CR confinement along magnetic field lines are highly uncertain. Modelling environment-dependent transport properties in galaxy simulations is still hazardous, so we do not attempt a self-consistent
calculation of the diffusion coefficient yet and we start by adopting simple isotropic and anisotropic diffusion in addition to gas advection.

Due to computational limitations in the numerical evolution of the cold gas, cosmological galaxy simulations have often limited the amount of cold gas allowed to exist, or used simplified thermal functions to avoid costly calculations. 
A common assumption is to consider collisional ionization equilibrium in which case the approach described by \cite{Sutherland1993} is sufficient to describe low-density plasma. However, in the presence of UV background, the assumption of collisional ionization equilibrium no longer holds \citep{Shen2010,Smith2017}. A more local approach has also been addressed in order to describe self-shielding regions when necessary, either with the assumption of equilibrium \citep{Gnedin2012} or without it \citep{Katz2022}. Other galaxy simulations only use an effective equation of state for the ISM, with radiative cooling and star formation using a probabilistic approach \citep{Springel2003}. Gas cooling and heating, as well as radiative transfer, are described in detail in the CR-inclusive FIRE-2 simulations as they aim to account for mechanical and radiative stellar feedback in galaxies \citep{Hopkins20,Hopkins2021}. 
Other CR-inclusive MHD simulations have been coupled to sophisticated chemical networks to accurately describe the gas phases, but they have been restricted to the modelling of a few specific starburst galaxies \citep{Krumholz20}, to outflow studies in a $0.5\times0.5\times\pm10$~kpc piece of galactic disc under solar-neighbourhood conditions \citep{Girichidis18}, and to ISM stratification studies as a function of SN frequency and SN location in a $1\times1\times\pm5$~kpc patch \citep{Simpson22}.

In order to retain enough computing flexibility to study the CR/ISM coupling for different CR transport modes in dwarf galaxies, then in massive galaxies (for their longer-lived spiral arms and disc stratification), while adequately modelling the gas circulation between phases, we have exploited the results of sub-parsec resolution simulations of individual cloud complexes. The latter account for detailed cooling and heating rates with computationally simple and efficient implementations \citep{Audit2005} that, while not accounting for the densest molecular clumps, successfully reproduce observed ISM features in the phases we seek to study \citep{Saury2014}. 

This paper is the first in a series where we address CR diffusion and the multi-phase evolution of the ISM in 9-pc-resolution simulations of a gas-rich dwarf galaxy in a $10^{11}$~M$_{\odot}$ halo. We focus here on the steady-state, large-scale properties of the galaxy, and defer the discussion of the gas and CR structures at sub-galactic scales to the next article. This paper is structured as follows: in section \ref{sec:SimMethods} we describe the methods used to perform the simulations; in section \ref{sec:Results} we present and discuss the global properties of the simulated galaxies, focusing on galactic gradients in section \ref{sec:gradients}, on mass fractions in the different gas phases in section \ref{sec:ISMphases}, and on the spatial distribution of the gas phases in section \ref{sec:gradphase}. We present the results on  star formation in section \ref{sec:SFR}, estimating the global and local rates in section \ref{sec:SFRandKS} and discussing the CR feedback on star formation in section \ref{sec:CRandSFR}. We present the \g-ray results in section \ref{sec:gamma}, discussing how the simulated galaxies compare with the updated version of the \g-ray versus SFR relation in section \ref{sec:gammaSFR} and comparing them with other simulations in section \ref{sec:simcomp}. Finally we present our conclusions in section \ref{sec:Conclusions}.
\section{Simulation methods}\label{sec:SimMethods}
We use the adaptive mesh refinement code \textsc{RAMSES} \citep{Teyssier2002} to compute the magneto-hydrodynamic (MHD) evolution of an idealized, isolated dwarf galaxy following the method described by \cite{Dashyan20}. This simulation includes stars and gas inside a dark matter halo. The gas dynamics follow the equations of ideal MHD evolution \citep[as in][]{Fromang2006}, computed with Harten-Lax-van Lear Discontinuities Riemann solver from \cite{Miyoshi2005}

To allow comparison with the results of \cite{Dashyan20}, we have used  
the same initial conditions for the galaxy labelled G9, the main parameters of which are listed in their Table 1. It includes a $10^{11}$~M$_{\odot}$ dark-matter halo and a baryonic mass of $\sim 4 \times 10^9$~M$_{\odot}$.
These initial conditions where first used by \cite{Rosdahl2015,Rosdahl2017} and generated with the \textsc{makedisc} code \citep{Springel2005}. It consist of a disc galaxy with a scale radius of 1.5~kpc and a scale height of 150~pc. It is evolved inside a 150~kpc wide cubic box with zero-gradient boundary conditions on all sides. In the simulation, the lower and higher resolution volume elements correspond to refinement levels of 7 and 14, meaning cubes of $\sim$1172~pc and $\sim$9~pc of side respectively. The initial magnetic field is toroidal in the galactic plane. Its strength scales with the gas density as $B \propto \rho^{2/3}$ to mimic the effect of pure compression of magnetic-field lines in the disc. It is set at a value of 1 $\mu$G at a gas density of 15 cm$^{-3}$ \citep[see eq. 1 of][]{Dashyan20}. 

These simulations used gas heating and cooling prescriptions summarised below. To reduce possible fragmentation of dense clouds at the smallest scales \citep{Truelove1997}, \cite{Dashyan20} imposed a minimum pressure (i.e. a "pressure floor") that scales quadratically with the gas density to follow the evolution of the Jean's mass \citep{Teyssier10}.
This floor drastically limits the ISM evolution at the scales of tens and hundreds of parsecs we are interested in to study CR feedback on the ISM. We have therefore improved the gas thermal evolution in the simulation setup by implementing heating and cooling functions that allow the thermal instability to partition the gas between the warm and cold neutral phases of the ISM (see section \ref{sec:cooling} for details). 


\begin{figure}[ht!]
   \centering

      \includegraphics[width=0.49\textwidth]{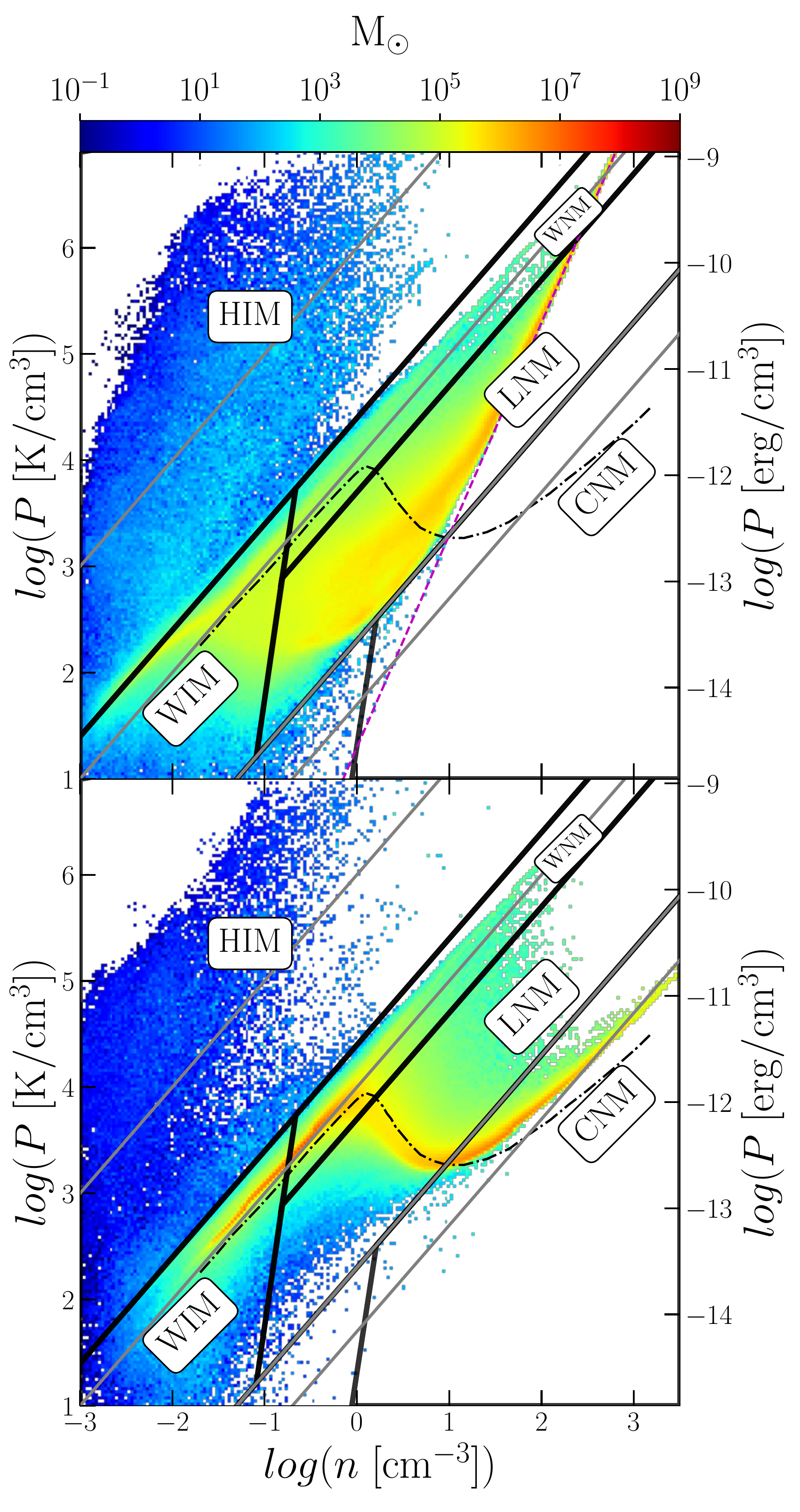}
      
      \vspace{-3\baselineskip}
      \vspace{2\baselineskip}
      \caption{Phase diagrams of the same simulation, performed on the top panel with the pressure floor (dashed magenta line) and cooling functions of \cite{Dashyan20}, and on the bottom panel, with the bimodal distribution of the neutral gas in the warm and cold phases. Solid black boundaries are given to partition the gas phases between the warm and cold neutral media (WNM and CNM, respectively), the unstable lukewarm neutral medium (LNM), and the warm and hot ionised media (resp. WIM and HIM). The dot-dashed curve corresponds to the gas in thermal equilibrium according to \cite{Wolfire1995}. Isothermal lines at 50, 200, $10^4$ and $10^6$ K are shown in grey.}
         \label{Fig:PhasediagPflorNoPflor}
\end{figure}

\subsection{Star formation and feedback}
\label{sec:feedback}
Star formation and SN feedback prescriptions follow those of the \cite{Dashyan20} runs. The star-formation model is based on the Kennicutt-Schmidt law \citep{Kennicutt1998,Krumholz2007} with a fixed star-formation efficiency, $\epsilon_{*} = 0.02$, that dictates the fraction of gas density, $\rho$, that is turned into stars per free-fall time, $t_{\textrm{ff}}$, i.e. $\dot{\rho}_* = \epsilon_{*} \rho/t_{\textrm{ff}}$, in a cell that is tagged as "star forming". Star-forming cells are those colder than $3\times 10^3$ K and denser than the number density threshold, $n_0 = 80\, \rm H\,$~ cm$^{-3}$. Star formation is a stochastic process drawn from a Poisson distribution \citep{Rasera2006} that results in particles with stellar mass of $m_{*} = 2 \times 10^3$  M$_{\odot}$.

Given the stellar initial mass function of \cite{Chabrier2003}, a fraction $\eta_{\textrm{SNe}}$ of this stellar mass will end as core-collapse SNe. 
The SN feedback is triggered at $t_{\textrm{SN}} = 5$ Myr after the formation of a star particle. It involves the injection of a total energy
\begin{equation}
E_{\textrm{SNe}} = 10^{51} \eta_{\textrm{SNe}} \frac{m_*}{m_{\textrm{SN}}} \;\textrm{erg},
\end{equation}
 \noindent and a mass of ejecta,
 $m_{\textrm{ej}} = \eta_{\textrm{SNe}} m_*$,
 \noindent where $m_{\textrm{SN}}$ is the typical mass of a type II SN progenitor. Values of $\eta_{\textrm{SNe}}=0.2$ and $m_{\textrm{SN}}=20$ M$_{\odot}$ are inferred from the \cite{Chabrier2003} stellar initial mass function. 
 
 For each SN event, a fraction $\eta_{\textrm{CR}}=10$\% of the total SN energy is injected in the form of CRs in the supernovae neighbourhood (see \cite{Dashyan20} for details on CR injection). The remaining energy is injected thermally following the mechanical feedback developed by \cite{Kimm2014} and \cite{Kimm2015}. It is distributed among the neighbouring cells depending on the gas mass and momentum in them in order to reproduce the two main stages of supernova evolution, the adiabatic phase and the momentum conserving or "snow-plough" phase. 
 
 We note that, while CRs are likely accelerated by SN shock waves \citep{Blasi13,Bykov18,Gabici19}, the $\eta_{\textrm{CR}}$ fraction of SN kinetic energy that is imparted to CRs is rather uncertain. Early acceleration models suggested a 10\% conversion efficiency \citep{Markiewicz90}. Improved acceleration models including the CR back-reaction on the shock MHD structure suggest values as high as 50\% \citep{Ellison04} whereas PIC simulations limit this efficiency to be 10-20\% when the magnetic field is nearly aligned with the shock velocity, and much less in more oblique configurations \citep{Caprioli2014}, so the average efficiency of a SN remnant depends on the magnetic coherence length in its environment \cite{Dubois19,Pais20}. \g-ray observations of a sample of SN remnants point to values well below 10\% in certain objects \citep{Acero16SNR} and in the 25\%-50\% range in others \citep{Ellison12}.
 An average efficiency around 5\%-10\% is derived from the global \g-ray intensity of the Milky Way \citep{Strong2010}, but the contribution of core-collapse supernovae to ten per cent of the observed IceCube neutrino flux above 60~TeV necessitates that 20\%-30\% of the shock energy be channelled to CRs \citep{Petropoulou17}. Given these uncertainties, we have adopted the canonical $\eta_{\textrm{CR}}=10$\% conversion efficiency and we defer to subsequent work the study of a broader range of efficiencies. 

\subsection{Gas evolution} \label{sec:cooling}
The simulations produced by \cite{Dashyan20} used the  metal-dependent cooling function for gas hotter than $10^4$ K from \citep{Sutherland1993}, valid for low-density plasma at collisional ionization equilibrium. For gas below $10^4$ K, they followed the approximations for radiative cooling used in \cite{Rosen1995} from \cite{Dalgarno1972} and \cite{Raymond1976}. Regarding the gas heating, aside from the SN injection, they accounted for the ultraviolet background heating for diffuse gas \citep{Haardt1996}. Compared to this setup, the two main updates in the present runs are the removal of the pressure floor and the implementation of heating and cooling functions that allow cold clouds to form out of the warm neutral gas by the well known thermal instability \citep{Field1969}.

The idea that two neutral components of the ISM can exist simultaneously at thermal equilibrium and at different thermodynamical states, warm-diffuse and cold-dense, is supported both theoretically \citep{Field1969,Mckee1977,Wolfire1995} and observationally \citep[e.g.][]{Low1984,Heiles03,Murray2018}. To reproduce this bi-modal state, we use a set of cooling and heating functions that has been implemented by \cite{Audit2005}, that is pictured in Fig. 1 of \cite{Saury2014}, and that has been tested against ISM data in sub-pc resolution simulations \citep{Audit10,Saury2014}. The dominant cooling is due to CII, OI, and Lyman $\alpha$ line emissions, and to electron recombination onto positively charged grains. Heating is dominated by the photo-electric effect on small grains and polyaromatic hydrocarbons due to far-ultraviolet radiation. A spatially uniform radiation field with the spectrum and intensity of the Habing field ($G_0/1.7$ where $G_0$ is the Draine flux) \citep{Habing68,Draine78} is assumed in the simulation. The heating from soft X-rays is ignored as it is thought to be negligible and the heating from CRs is implemented through the CR pressure in the MHD solver \citep{Dashyan20}. The ionisation in the cooling functions and in the gas phase characterisation (see section \ref{sec:ISMphases}) is calculated using the approximation proposed by \cite{Wolfire2003}. Regarding metallicity, the cooling module is based on local abundances in the solar neighbourhood.

For the thermal instability to develop, the warm neutral gas must be dynamically perturbed at a scale greater than the sound crossing scale which is the sound speed times the cooling time \citep{Saury2014}. It corresponds to $30$~pc for the typical $10^4$~K and 0.5~cm$^{-3}$ conditions of the warm neutral gas. This scale is resolved in our 9-pc simulations.   


To illustrate the effects that each gas physics set-up has on the ISM, Fig. \ref{Fig:PhasediagPflorNoPflor} shows the phase diagrams of the same simulation performed with the cooling functions and pressure floor of \cite{Dashyan20} on the top panel, and with the new cooling and heating functions on the bottom panel.  The dot-dashed curve shows the gas at thermal equilibrium \citep{Wolfire1995}. The equilibrium is stable only in the region where $dP/d\rho$ is positive and the gas distribution does show a relative depletion in the unstable part of the equilibrium ($dP/d\rho < 0$). We note that the best resolution (9~pc) in the simulation does not allow to follow the coldest and densest part of the distribution that would further develop along the Wolfire curve and transition to \hd at higher resolution. The pressure floor used by \cite{Dashyan20} and depicted by the dashed magenta line drastically affects the gas distribution, preventing the gas from condensing to the cold stable phase and keeping it into the region of thermal instability (labelled as LNM see section \ref{sec:ISMphases}), or pushing it to the warm and hot ionised phases (labelled WIM and HIM). In this set-up, the unstable phase is overly populated and it gathers the largest fraction of the gas mass, at variance with what is observed in the real ISM.

\subsection{Cosmic-ray diffusion}

   \begin{figure*}
   \centering
   \begin{subfigure}{0.89\textwidth}
      \includegraphics[width=\textwidth]{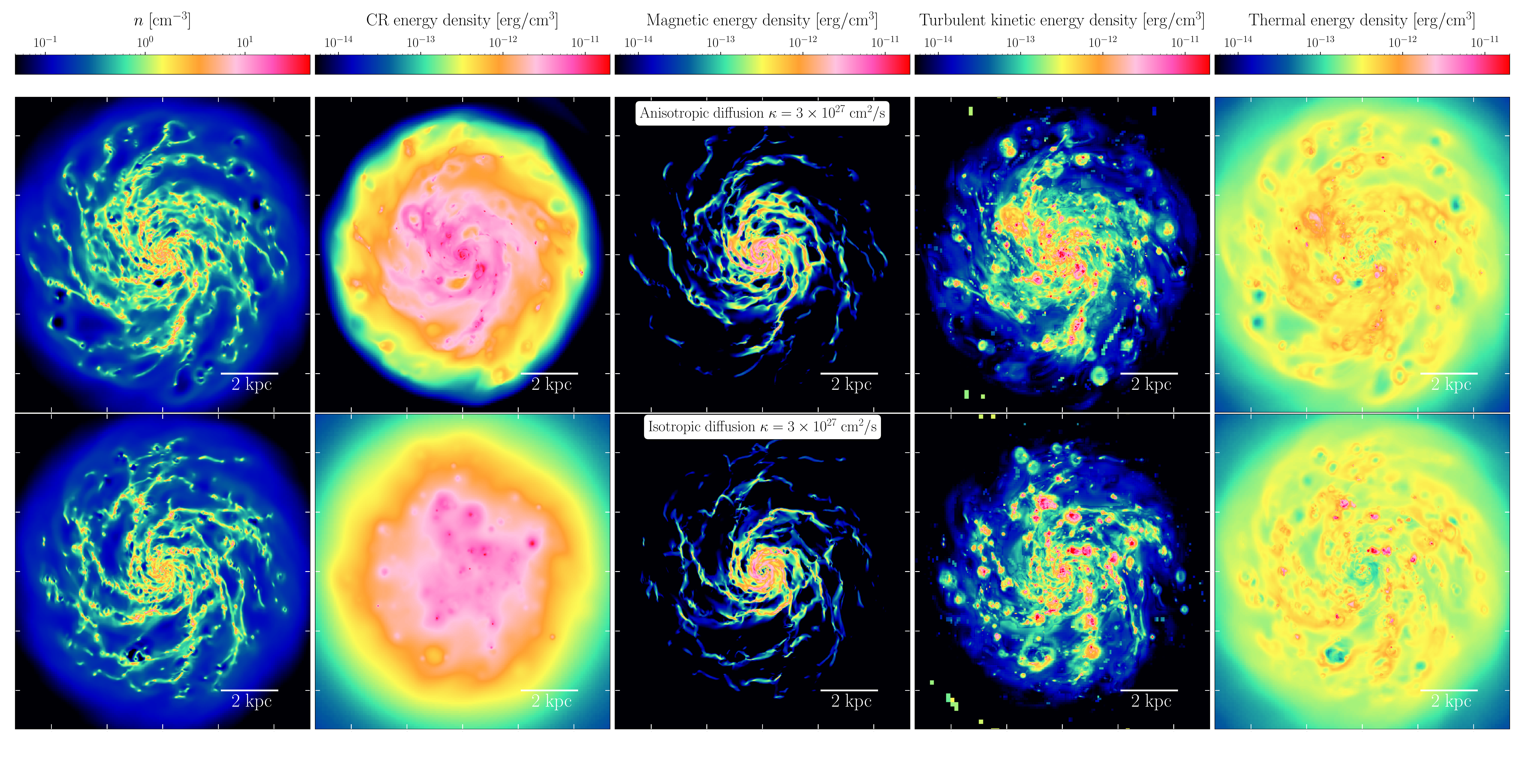}

   \end{subfigure}\\[-4ex]
   \begin{subfigure}{0.89\textwidth}
      \includegraphics[width=\textwidth]{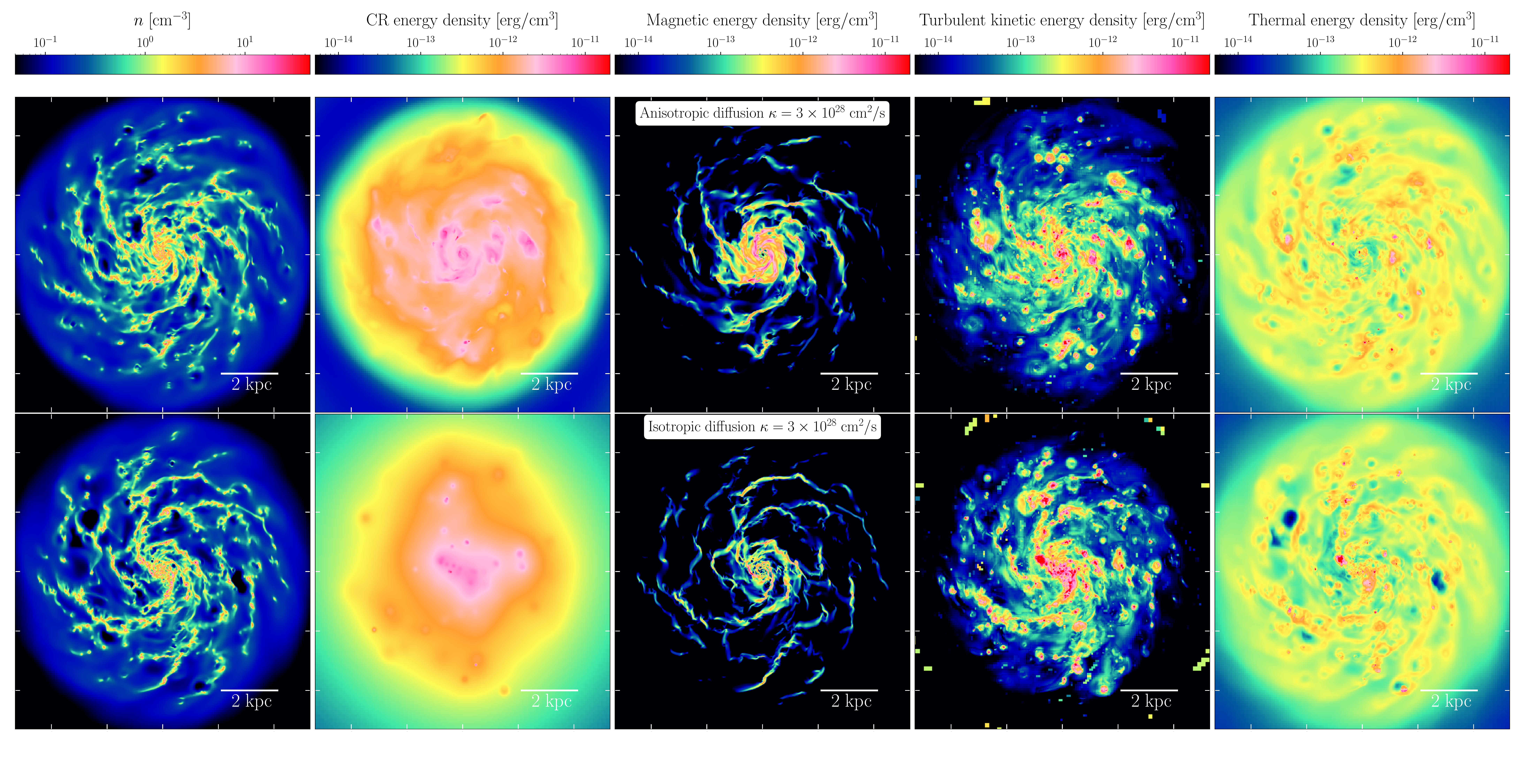}

   \end{subfigure}\\[-4ex]
   \begin{subfigure}{0.89\textwidth}
      \includegraphics[width=\textwidth]{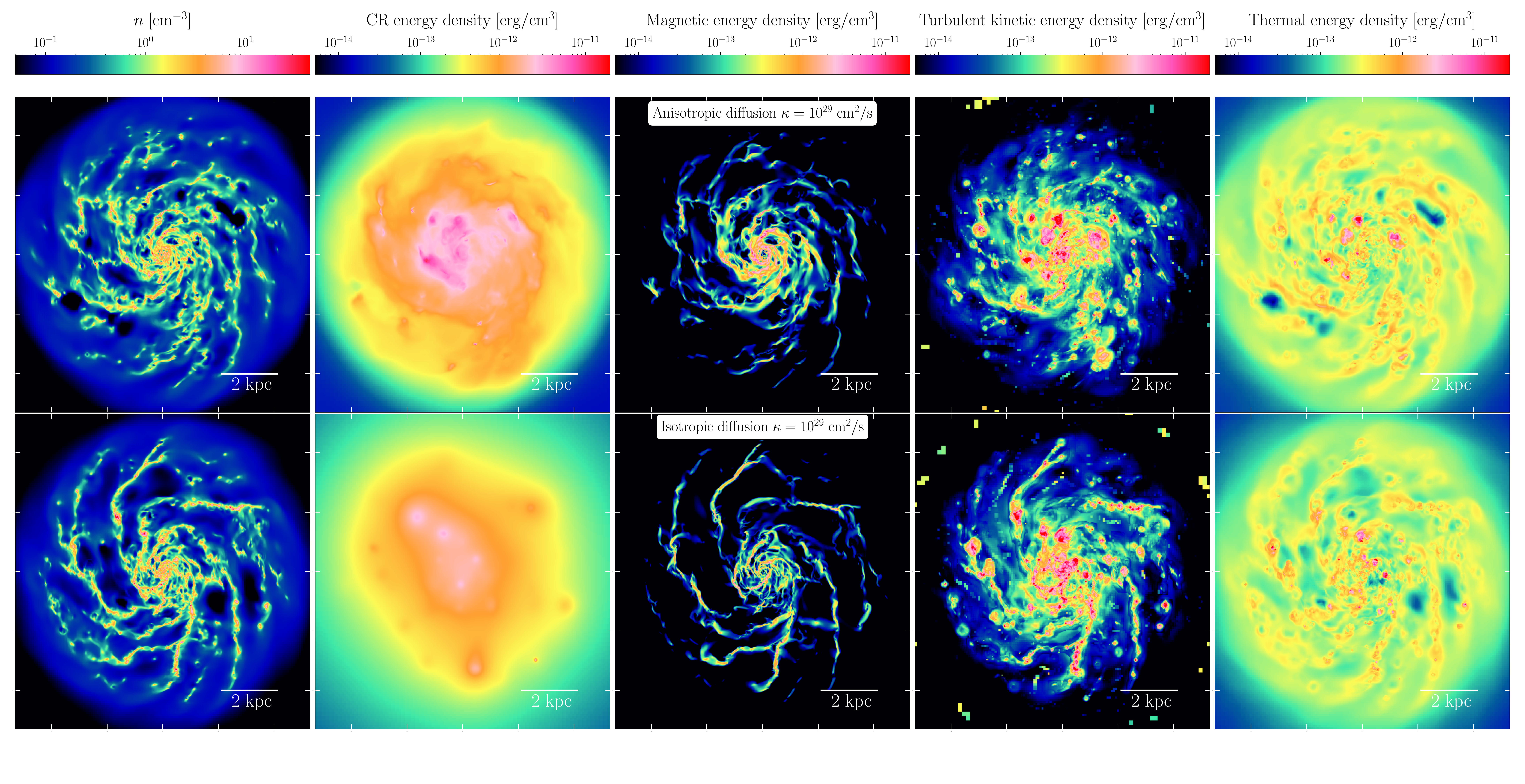}

   \end{subfigure}\\[-4ex]

      \vspace{-0.49\baselineskip}
      \caption{Face-on maps from left to right of the gas density and of energy densities in CRs, magnetic field, turbulent gas motions, and thermal. The maps are grouped by CR diffusion coefficient to compare the isotropic and anisotropic cases. All energy densities share the same colour scale.}
         \label{Fig:EnergyMapsFaceon}
   \end{figure*}
   \begin{figure*}
   \centering
   \begin{subfigure}{0.89\textwidth}
      \includegraphics[width=\textwidth]{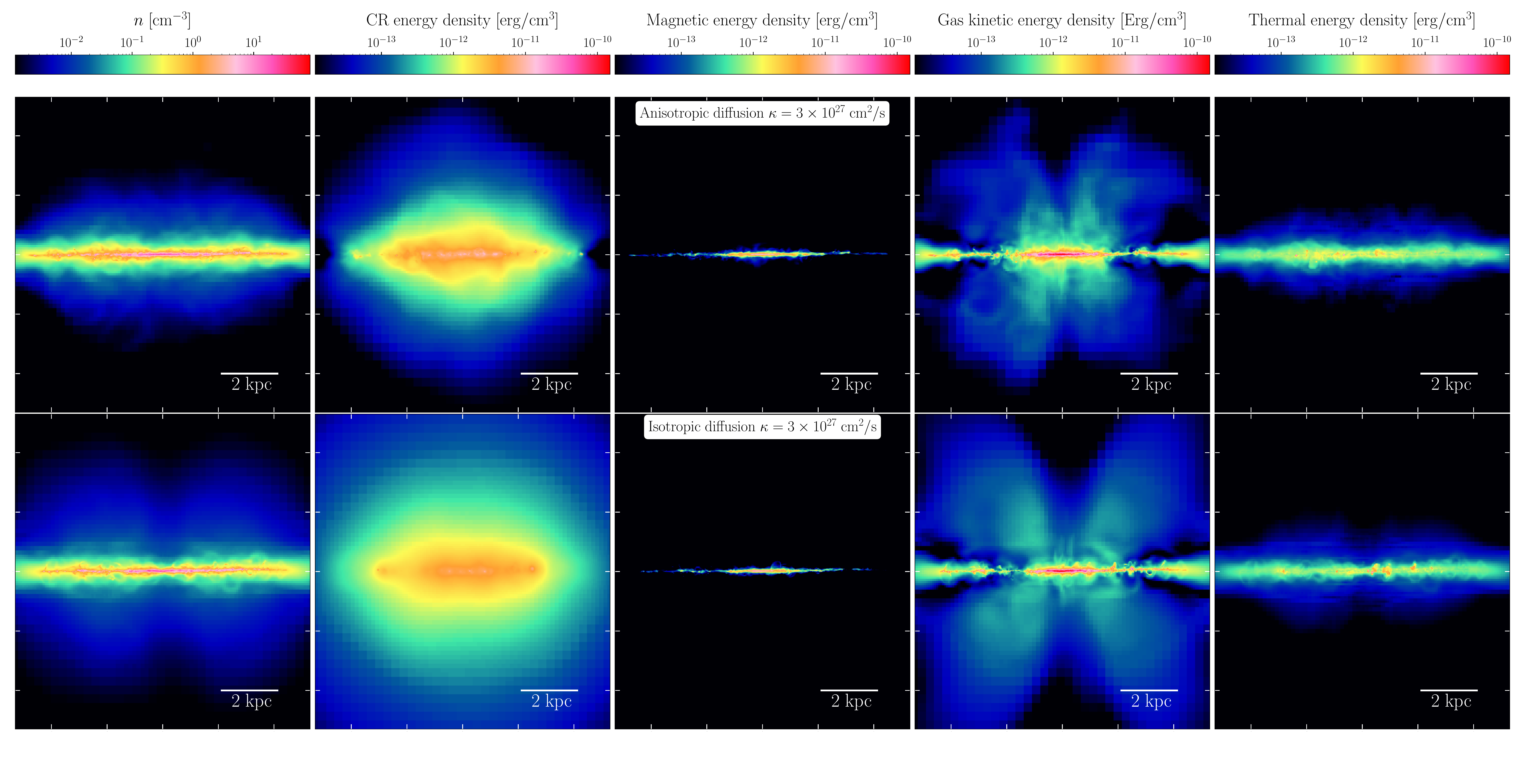}
   \end{subfigure}\\[-4ex]

   \begin{subfigure}{0.89\textwidth}
      \includegraphics[width=\textwidth]{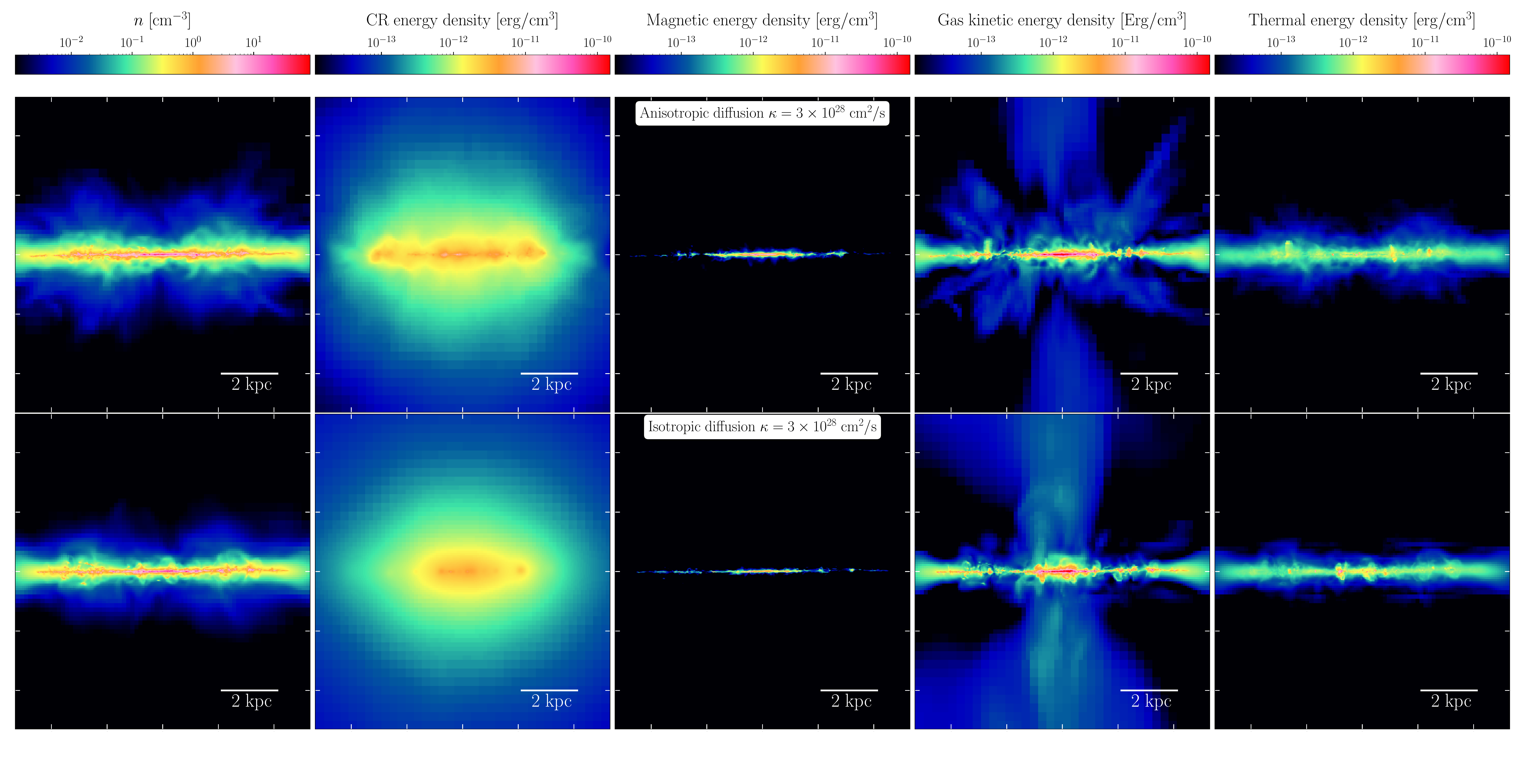}
   \end{subfigure}\\[-4ex]
 
   \begin{subfigure}{0.89\textwidth}
      \includegraphics[width=\textwidth]{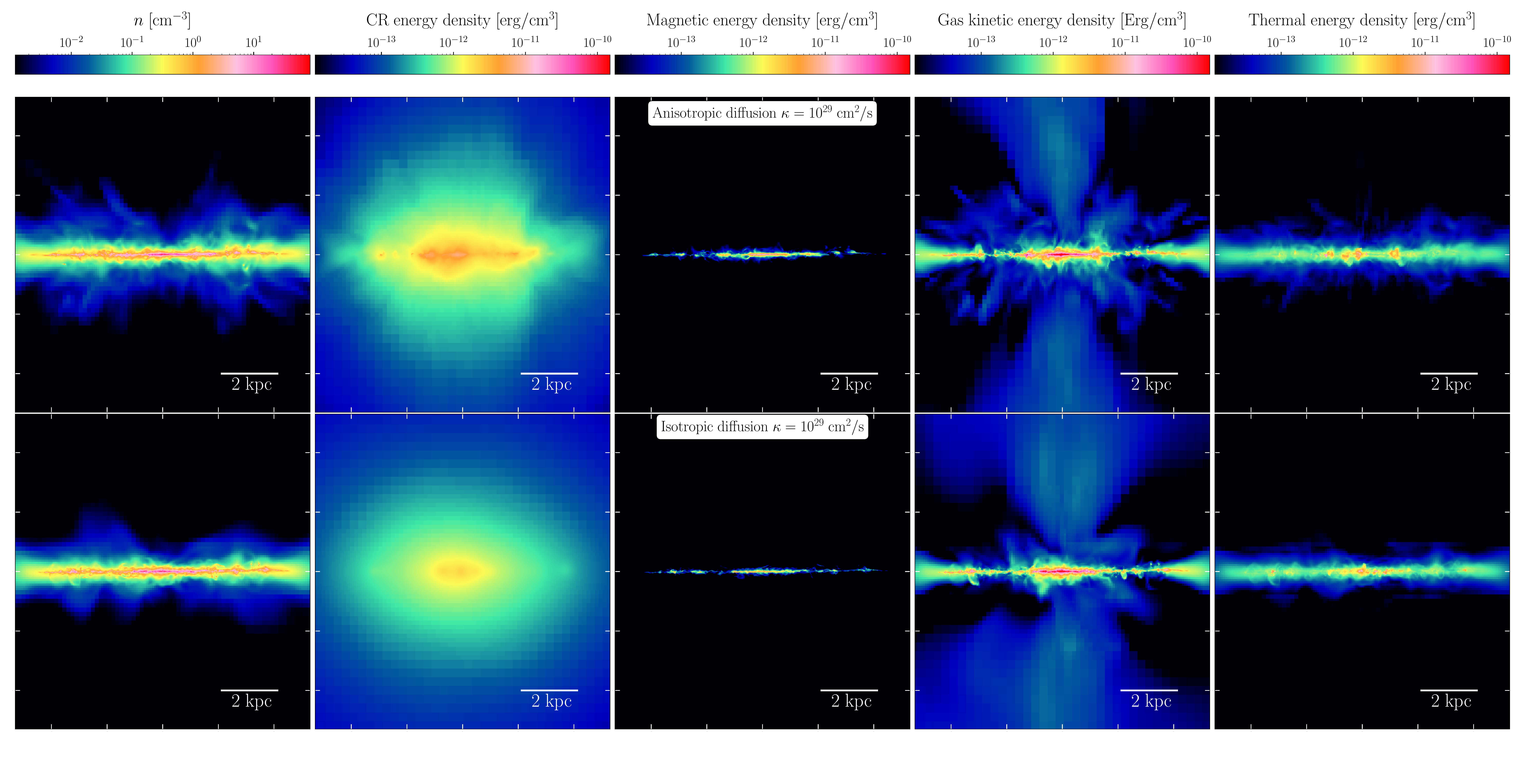}
   \end{subfigure}\\[-4ex]

      \vspace{2\baselineskip}
      \caption{Same as Figure \ref{Fig:EnergyMapsFaceon} for edge-on views, except for the turbulent kinetic energy which is replaced here by the kinetic energy density of the gas in the rotational frame of the galaxy.}
      \label{Fig:EnergyMapsEdgeon}
   \end{figure*}

The CR fluid emulates the evolution of low-energy (few GeV) CRs because they dominate the CR spectrum and energy density. The particles can be modelled as a fluid that exchanges energy and pressure with the thermal gas on macroscopic scales because their interaction with the gas is mostly collisionless and mediated by the magnetic field. The CR energy density is added to the kinetic, thermal, and magnetic ones in the total energy density in the ideal MHD equations: 
\begin{equation}
e_{tot} = \frac{1}{2} \rho v^2 + e_{th} +  \frac{B^2}{8\pi} + e_{CR}. 
\end{equation}
\noindent Similarly the total pressure accounts for the CR, thermal, and magnetic pressures: 
\begin{equation}
    P_{\textrm{tot}} = P_{\textrm{CR}} +P_{\textrm{th}} + \frac{B^2}{8\pi}
     = e_{\textrm{CR}}(\gamma_{\textrm{CR}}-1) + e_{\textrm{th}}(\gamma-1) + \frac{B^2}{8\pi}
\end{equation}
\noindent with adiabatic indices $\gamma_{\textrm{CR}}=4/3$ for a fully relativistic CR fluid and $\gamma=5/3$ for a monoatomic ideal thermal gas. 

To solve CR diffusion, we have used a combination of implicit and explicit finite-volume methods in the \textsc{RAMSES} code similar to that described in \cite{Dashyan20} and in \cite{Dubois2016}. We have implemented a switch, based purely on computing performance, to choose whether to calculate CR diffusion implicitly or explicitly depending on the ratio of the explicit diffusion step ($\textrm{d}t_{\textrm{dif}}=\Delta x^2/\kappa$) and the explicit hydrodynamic step ($\textrm{d}t_{\textrm{hyd}}=\Delta x/v$), where $\kappa$ notes the CR diffusion parameter, $v$ the gas velocity, and $\Delta x$ the cell size. We switch to the explicit diffusion solver (with sub-cycling) in levels where $\textrm{d}t_{\textrm{dif}}>\textrm{d}t_{\textrm{hyd}}/30$, and to the implicit method otherwise. 

Given the uncertainties we discussed in the introduction about the actual transport properties of CRs in the different interstellar environments and our interest in the gas structure in the disc, we have concentrated first on the simplest assumption of uniform diffusion, either isotropically or preferentially along the magnetic field. The latter choice allows to test the impact of a directed flow of CRs whereas isotropic diffusion likely applies in the neutral gas phases where the streaming instability is easily suppressed, but where a large fraction of the \g-ray luminosity is produced. MHD turbulence in these phases being often trans-\alfvenic or slightly super-\alfvenic \citep{Hu19}, the magnetic field is tangled and CR transport resembles diffusion due to field line random walk on scales larger than the coherence length of the field lines \citep{Xu22}. This length, $L_A = L_{inj} \mathcal{M}_{\rm A}^{-3}$ is around 1-10~pc for an injection scale $L_{inj}$ of order 100~pc and an \alfven Mach number, $\mathcal{M}_{\rm A}$, in the range of 2 to 5 (in appendix \ref{App:Machnumbers} we show maps of $\mathcal{M}_{\rm A}$ for our simulations). The \alfven Mach number, $\mathcal{M}_{\rm A}=v_{turb}/v_A$, compares the turbulent velocity of the gas, $v_{\rm turb}$, to the \alfven velocity, $v_A$. 
As noted by previous authors \citep[e.g.][]{Chan19,Dashyan20,Hopkins21gam,Bustard21}, adding streaming to diffusion has a small impact outside dense neutral clouds as the \alfven velocity hardly competes with diffusion in the more diffuse ISM: for a CR gradient scale length of order 1~kpc, streaming at a velocity $v_{st}$ comparable to the \alfven one corresponds to an effective diffusion coefficient of $3\times 10^{27}\, \rm (v_{st}/10\, \mathrm{ km\,s^{-1}}) (L_{CR}/1\,\mathrm{kpc})$~\cmsqs. 

We have tested three uniform diffusion coefficients, $\kappa =  3\times 10^{27}$, $3\times 10^{28}$, and $10^{29}$~\cmsqs, which bracket the typical value of $3 \times 10^{28} (R_{CR}/1\,{\rm GV})^{0.34}$~\cmsqs inferred as a function of particle rigidity $R_{CR}$ in the Milky Way \citep{Johannesson19}. They also correspond to the range of values inferred by \cite{Armillotta2022} in different interstellar environments.

In the case of anisotropic diffusion, the $\kappa$ values correspond to the diffusion coefficient along the magnetic field lines, and the perpendicular coefficient is a hundred times smaller: $\kappa = \kappa_{\parallel} = 100 \, \kappa_{\perp}$. This choice is motivated by the fact that, in mildly sub-\alfvenic turbulence \citep{Hu19,Hopkins21gam}, the perpendicular coefficient is smaller than the parallel one by a factor $\mathcal{M}_{\rm A}^4$ \citep{Yan08}.
We leave the study of spatially variable transport properties for the next simulations.


\begin{figure*}
   \centering
   \begin{subfigure}{0.95\textwidth}
      \includegraphics[width=\textwidth]{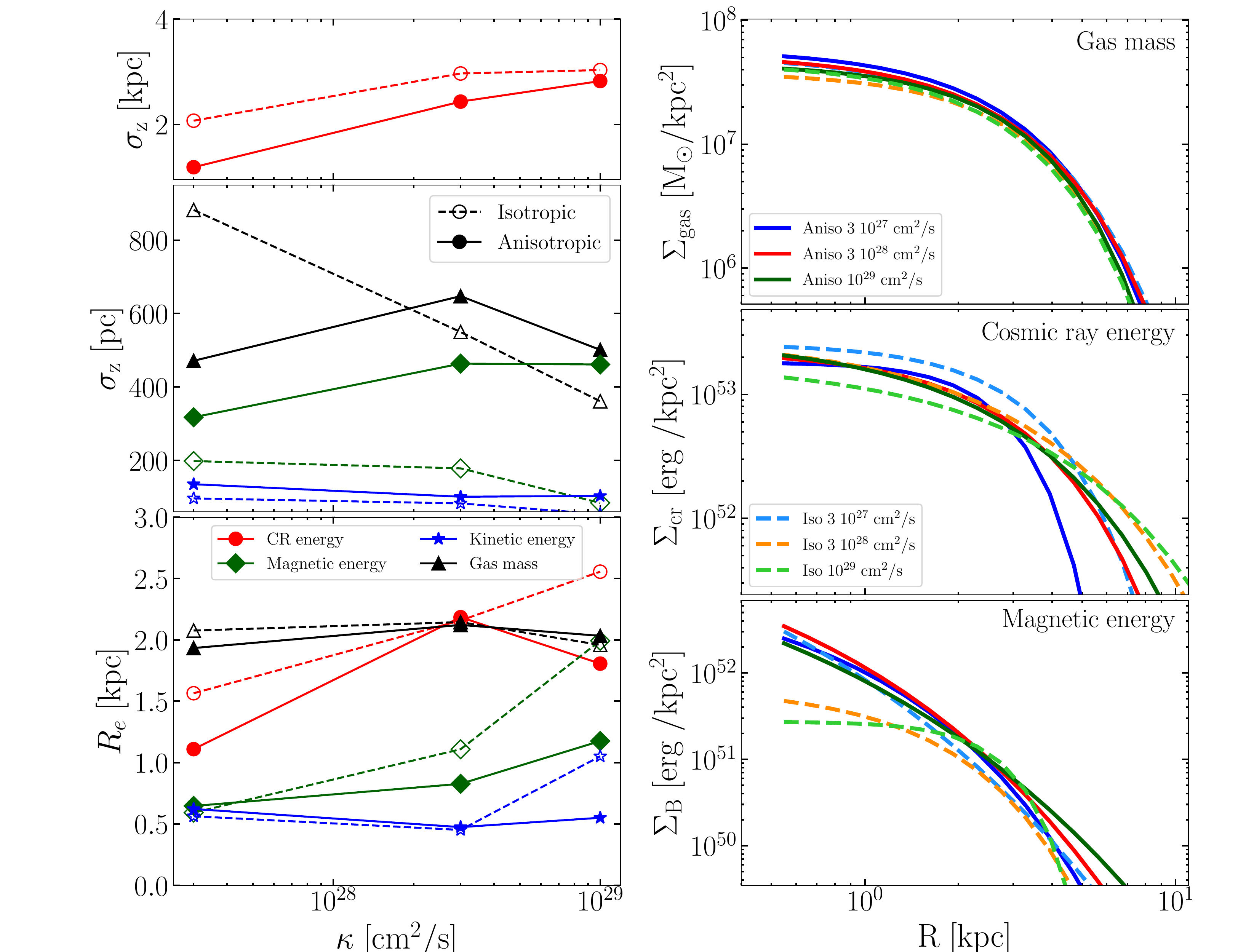}
      \vspace{-1.5\baselineskip}
   \end{subfigure}

      \vspace{2\baselineskip}
      \caption{Left: evolution as a function of the diffusion coefficient, $\kappa$, of the vertical scale height, $\sigma_{\textrm{z}}$ (top plot in kpc and middle plot in pc), and of the exponential decline radius, $R_{e}$ (bottom plot), for the gas mass and for different energy components as labelled in the bottom plot. Solid and dashed lines connect the values obtained with anisotropic and isotropic CR transport, respectively. Right: S\'ersic fits to the surface density profiles in gas mass (top), CR energy (middle), and magnetic energy (bottom) as a function of radial distance in the disc. The fits were obtained within $\pm 1 \sigma_z$ about the galactic plane.}
        \label{Fig:scaledistances}
\end{figure*}
\begin{figure*}
   \centering
   \begin{subfigure}{0.99\textwidth}
      \includegraphics[width=\textwidth]{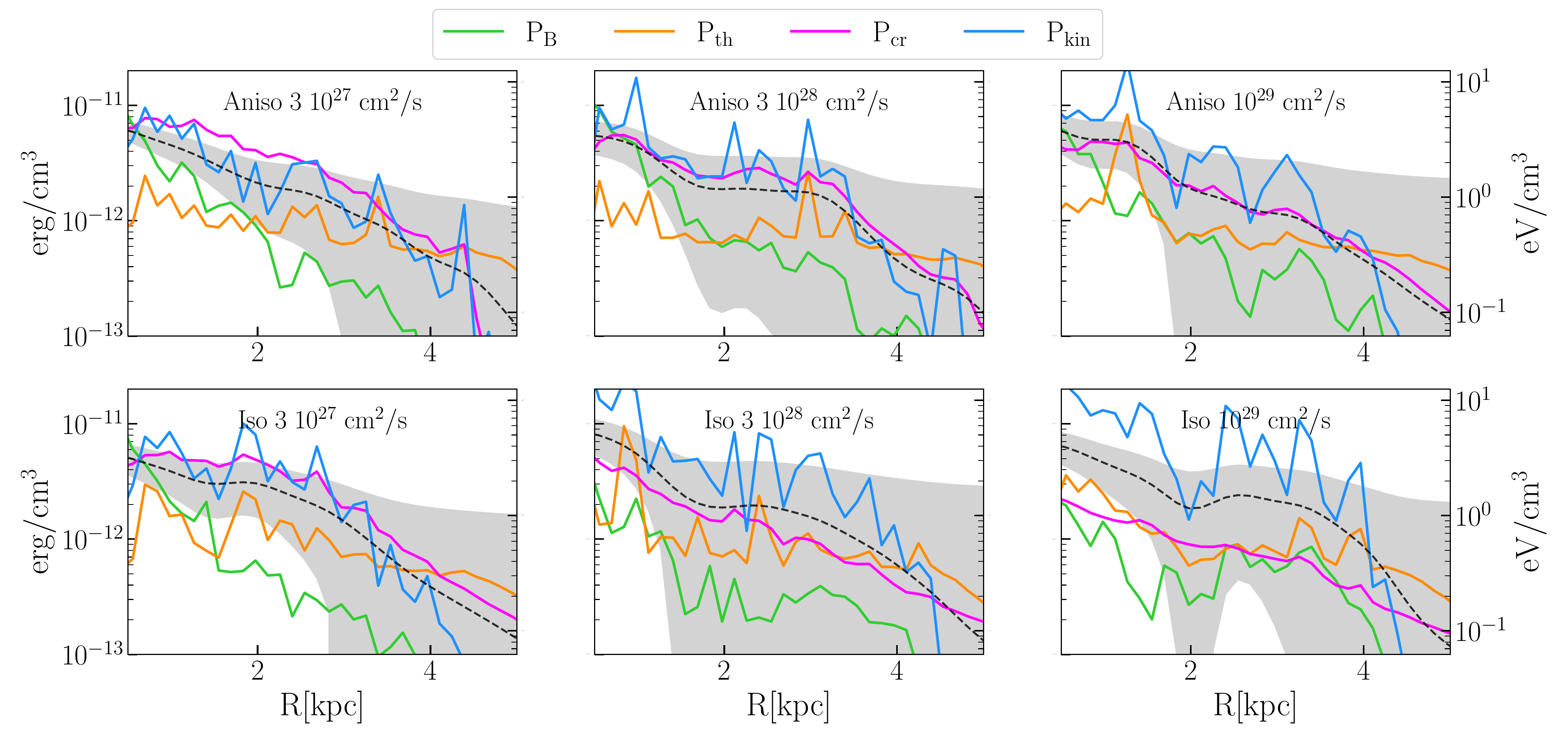}
      \vspace{-1.5\baselineskip}
   \end{subfigure}

      \vspace{2\baselineskip}
      \caption{Radial profiles of the magnetic (green), thermal (gold), CR (magenta), and kinetic (blue) pressures within $\pm$~200 pc from the galactic plane to capture mostly the neutral gas. The black dashed line shows the smoothed equipartition pressure and the grey band its rms dispersion given the radial fluctuations in each of the term. The kinetic energy is calculated form the turbulent RMS velocity as $P_{\rm kin}= \frac{1}{2} \rho \,\sigma_{turb}^2$.}
        \label{Fig:Pratios}
\end{figure*}
\section{Results on large-scale galactic properties}
\label{sec:Results}
We present the large-scale properties of our six realizations of the G9 galaxy, first focusing on the radial and vertical profiles of the different energy components and how CR transport affects them, then discussing the same distributions in the different ISM phases, before presenting how the overall star-formation rate and \g-ray luminosity vary with CR transport.  
In order to check that the galaxies are in an overall steady-state, the simulations have been run to several tens of Myr after the age of 250 Myr which we have adopted to present their properties.
We note $R$ and $z$ the radial and vertical cylindrical coordinates in the galactic disc. 

\subsection{Large-scale galactic gradients}\label{sec:gradients}

Figures \ref{Fig:EnergyMapsFaceon} and \ref{Fig:EnergyMapsEdgeon} show face-on and edge-on maps of the gas number density and of the energy densities in the CRs, magnetic field, and gas kinetic energy in the local rotational frame of the galaxy (based on the rotation curve\footnote{The kinetic energy is computed by considering the energy once the circular velocity corresponding to the local radius is substracted from the cell velocity i.e: 
\begin{equation*}
    K=\frac{\rho_i}{2} \sum_i (u_{\rm gas,i}-V_{\rm circ,i}(r_i))^2
\end{equation*}
\noindent here, $u_{\rm gas,i}$, $r_i$ and 
$\rho_i$ are the speed, the radius and the density of the gas in the i-th cell of the mesh. $V_{circ}(r)= (GM_{\rm tot}(<r)/r)^{1/2}$.
})

We show average values obtained in 2D square pixels, 9 pc in side and with a thickness equal to the specific scale height of each component on each side of the galactic plane for the face-on views, and 1~kpc thick slices across the galaxy for the edge-on views. The scale heights of the different components are presented in Fig. \ref{Fig:scaledistances}. In Figures \ref{Fig:EnergyMapsFaceon} and \ref{Fig:EnergyMapsEdgeon}, the CR diffusion coefficient increases from top to bottom and the plots are grouped to compare the isotropic and anisotropic cases for the same diffusion coefficient. 

To complement the maps, we show the vertical scale height, $\sigma_z$, and the exponential decline radius, $R_e$, of each of the components in Fig. \ref{Fig:scaledistances}. 
The vertical rms dispersion is calculated from all cells lying at $1<R<6$ kpc and $|z|<10$ kpc as
\begin{equation}\label{eq:verticaldisp}
    \sigma_{z} = \sqrt{\frac{\sum_i w_i z_i^2}{\sum_i w_i} }.
\end{equation}
\noindent where $w_i$ corresponds to the gas mass or to the energy density of the component, and $z_i$ is the cell distance to the midplane.

The e-folding radius, $R_e$, is derived from the radial profile of the surface density, $\Sigma(R) \propto e^{-R/R_e}$, of the total gas mass or of the CR, magnetic, or kinetic energies. It is calculated from a linear fit of $\ln(\Sigma)$ versus $R$ in the 1 to 4~kpc range, and for a thickness on each side of the midplane that equals the specific scale height of each component.
Additionally, to look at the behaviour of the surface densities for either the gas mass or the magnetic and CR energies, we have fitted a S\'ersic profile to the corresponding data in concentric rings, each having a radial width of 1 kpc and a vertical thickness equal to the vertical scale height of the component. The resulting curves are shown in the right panels of Fig. \ref{Fig:scaledistances}. 

We have checked the time stability of these scale heights and radial profiles by selecting epochs exhibiting different star-formation rates (SFR), and epochs before and after star-formation bursts. Examples at different times are shown in appendix \ref{sec:time}. The profiles show that the galaxies have converged to realistic values, with ${\sim}$kpc-scale mean densities in the inner regions around 1~cm$^{-3}$ in gas and 0.8~eV~cm$^{-3}$ in CRs. Magnetic strengths reach a few microGauss in outer-galaxy filaments and values up to 10-20 $\mu$G in inner filaments, depending on CR transport. These values compare well with Milky Way measurements even though we simulate gas-rich dwarfs. In particular, the mean CR density matches the value of 0.9~eV~cm$^{-3}=1.4\times 10^{-12}$~erg~cm$^{-3}$ we derive for CR nuclei above 1 GeV/nucleon in the solar neighbourhood from the latest AMS-02 and Voyager measurements (see section \ref{sec:gamma}), or the density of 0.83-1.02~eV~cm$^{-3}$ obtained when including CR nuclei and electrons above 3 MeV \citep{Cummings16}. 

The face-on CR energy density maps (Fig.~\ref{Fig:EnergyMapsFaceon}) clearly illustrate the difference between the two diffusion scenarios, as the anisotropically diffusing CRs follow closely magnetised filaments whereas the isotropically diffusing ones gather in broad clumps around their sources. In both cases, CR structures get evidently washed out and the centre-to-edge gradient across the disc is shallower for faster diffusion. We note that these gas-rich dwarf galaxies have a SFR close to that of the Milky Way (see section \ref{Fig:SFR}) in a much smaller volume. This is why, in the inner star-forming regions, the CR energy density can locally exceed by a factor of 2 or 3 the solar value.

Comparing the concentration of CRs along the galactic plane in the edge-on maps of Fig. \ref{Fig:EnergyMapsEdgeon}, we see that large CR densities can extend to larger radii in the anisotropic case than in the isotropic one for the same diffusion coefficient because the particles flow along the spiral-like magnetic filaments rather than diverging out in all directions. Perpendicularly to the disc, the anisotropic CR scale height is smaller than the isotropic one for the same reason since the magnetic-field configuration is predominantly toroidal in the disc and is concentrated near the galactic plane.
As expected, the CR density shows shallower radial gradient and larger scale height off the plane for faster diffusion.

Except for a small difference in contrast between the densest and most diffuse regions and in the development of inter-arm spurs, the gas density distribution appears to be rather robust against changes in CR diffusion. The radial profile of the mass surface density in the disc is insensitive to CR changes (see Fig. \ref{Fig:scaledistances}). Even though we modify the diffusion coefficients by thirty, we do not detect any change in the gaseous disc size contrary to the findings of \cite{Buck2020} where multiplying the anisotropic diffusion coefficient by three (between $10^{28}$ and  $3\times10^{28}$~\cmsqs) resulted in a two to three times smaller disc radius (see the CRdiff case in their Fig. 15, which is close to our anisotropic case). Our galaxies have evolved in isolation whereas those of \cite{Buck2020} have evolved in a cosmological context. The different behaviours between the two sets support their conclusion that CR feedback impacts the disc size primarily by modifying the circulation of circumgalactic gas, hence the gas accretion history onto the galaxy.

The vertical distribution of the dense gas and of the magnetic field are rather insensitive to changes in diffusion coefficient in the anisotropic case because the particle flux is constrained by the toroidal magnetic field. In the isotropic case, on the contrary, the vertical scale heights of both the gas and magnetic field increase by a factor of two with decreasing $\kappa$ because the CR density builds up on shorter scales around the sources and it maintains the gas and the frozen-in magnetic field in hydrostatic equilibrium to larger heights above the plane.

Conversely, the magnetic energy density shows a strong radial gradient that significantly responds to the degree of anisotropy in CR transport. It is more centrally concentrated
for very slow and/or anisotropic diffusion.
Figure \ref{Fig:scaledistances} illustrates that the radial gradient is shallower (larger $R_e$) for isotropic diffusion and even more so as $\kappa$ increases. This behaviour qualitatively agrees with the results of \cite{Pakmor16} and \cite{Buck2020}, who have argued that the anisotropic diffusion of CRs allows for more turbulent fields and a stronger small-scale dynamo~\citep[see also][]{Hanasz09}.  
Additionally, one can note that, for faster diffusion and particularly so for the isotropic case, larger cavities develop in the ISM and the magnetic energy remains concentrated in sharper and fewer filamentary structures. We will characterise and discuss the degree of clumpiness and of filamentary contrast at sub-galactic scales in a companion paper.

The impact of CR diffusion on the central wind is clearly visible in Fig. \ref{Fig:EnergyMapsEdgeon}. For larger diffusion coefficients, the increased CR pressure above the disc helps to push the gas out. As the diffusion approximation we are using for CR transport is not appropriate for the diffuse ionised regions away from the disc, where self-excited \alfven waves should efficiently scatter the particles and make them stream at the local \alfven speed, we do not further discuss the wind and halo properties. 

\subsection{Large-scale pressure gradients}

Figure \ref{Fig:Pratios} shows the radial profile of the thermal, turbulent, magnetic, and CR pressures within ${\pm}200$~pc from the galactic plane. We compare them to the mean profile that corresponds to equipartition between the four terms. The pressure in turbulent gas motions is derived from the kinetic energy density in the turbulent RMS velocity field, $\sigma_{\rm turb}$, including the compressive and solenoidal components. The radius-to-radius fluctuations visible in Fig. \ref{Fig:Pratios} reflect the degree of clumpiness intercepted by the 150-pc-wide galactocentric rings for the different components (see Fig. \ref{Fig:EnergyMapsFaceon}). 

At large scales, the magnetic pressures are the weakest component except in the inner regions where they approach equipartition with the other terms when increased SFRs and larger CR pressures favour magnetic amplification. Otherwise, the magnetic pressures are typically a factor of five below equipartition. During the first half rotation period or so, all the galaxies form stars at the same rate and their average magnetic field grows similarly. It likely grows at these early stages from compression and turbulent dynamo at scales less than the turbulence driving scale. The growth slows down because of the back-reaction of the Lorentz force upon the turbulent motions \citep{Seta20}, but differential rotation, shear, and density stratification can further order and amplify magnetic fields on larger galactic scales. 
In the two cases where CR pressure feedback is weakest in the disc ($\kappa \geq 3\times 10^{28}$~\cmsqs isotropic diffusion), the growth of the average field stalls after the first rotation period and the large-scale magnetic pressures saturate at 5\%-10\% of the turbulent pressure throughout the disc in Fig. \ref{Fig:Pratios}.This saturation level compares well with the results obtained at higher resolution when forcing compressive or solenoidal turbulence in a multiphasic, resistive, and viscous medium \citep{Seta22}. For the other cases with much stronger CR feedback, in particular for all the anisotropic diffusion cases, star formation proceeds at a reduced rate in the inner regions (see section \ref{sec:SFR}), generating lower turbulent energy densities in these regions (see Fig. \ref{Fig:EnergyMapsFaceon} and \ref{Fig:Pratios}). Galactic outflows are also weaker and the average magnetic-field strength keeps growing with a doubling time close to the rotation period, as expected from CR-driven dynamo simulations \citep{Hanasz09}. The magnetic pressures reach up to 50\% of the turbulent pressures in the central regions and 20\% in the outer regions. We further discuss CR feedback on SFR and magnetic strength in section \ref{sec:CRandSFR}. 

{here some sentences and references about the timescales for magnetic growth. Are our dwarfs too young to have reached saturation? Arturo, what does the $<B_{CNM}>(t)$ curve say? probably not as we need about 10 rotations to build up B...}

Figure \ref{Fig:Pratios} shows that the turbulent and CR pressures often dominate the large-scale dynamics of the galaxy. The dominance of the turbulent component that is seen in the isotropic $\kappa \geq 3\times 10^{28}$~\cmsqs cases and in the anisotropic $10^{29}$~\cmsqs one is driven by larger SFRs which are presented in section \ref{sec:SFR}.
When CRs remain concentrated near their sources in star-forming regions, i.e. for very slow isotropic and/or anisotropic diffusion, both the CRs and the turbulence injected by SNe provide equivalent pressure contributions, regardless of the diffusion coefficient. When CRs are more diluted, either in the outer half of the galaxies, or because they rapidly diffuse to the halo, their pressure falls below the turbulent one, but remains comparable to the thermal values. 

We caution the reader, however, that the dominance of the turbulent and CR pressures over the thermal and magnetic ones should be further tested. Figure \ref{Fig:Pratios} shows that, outside the inner regions where CR-related magnetic amplification occurs, the profiles of the thermal and magnetic pressures are rather stable against two-fold changes in global SFR or against changes in CR transport. This is not the case for the turbulent and CR pressure profiles which respond to the SN rate for obvious reasons, via the total amount of energy and momentum injected into the gas by SN events, and by the fraction of SN energy imparted to CRs. We have taken standard values of $10^{51}$~erg for the energy released per supernova and 10\% for the energy fraction transferred to CRs, but variations around these values should be tested to judge the robustness of the turbulent and CR dominance over a broad range of anisotropic diffusion speeds. Since both these feedback parameters are subjects of extensive discussions, we recognize that it would be beneficial to further explore their effect on the dynamics of the ISM and the galaxy.  

\begin{figure}
   \centering
      \includegraphics[width=0.49\textwidth]{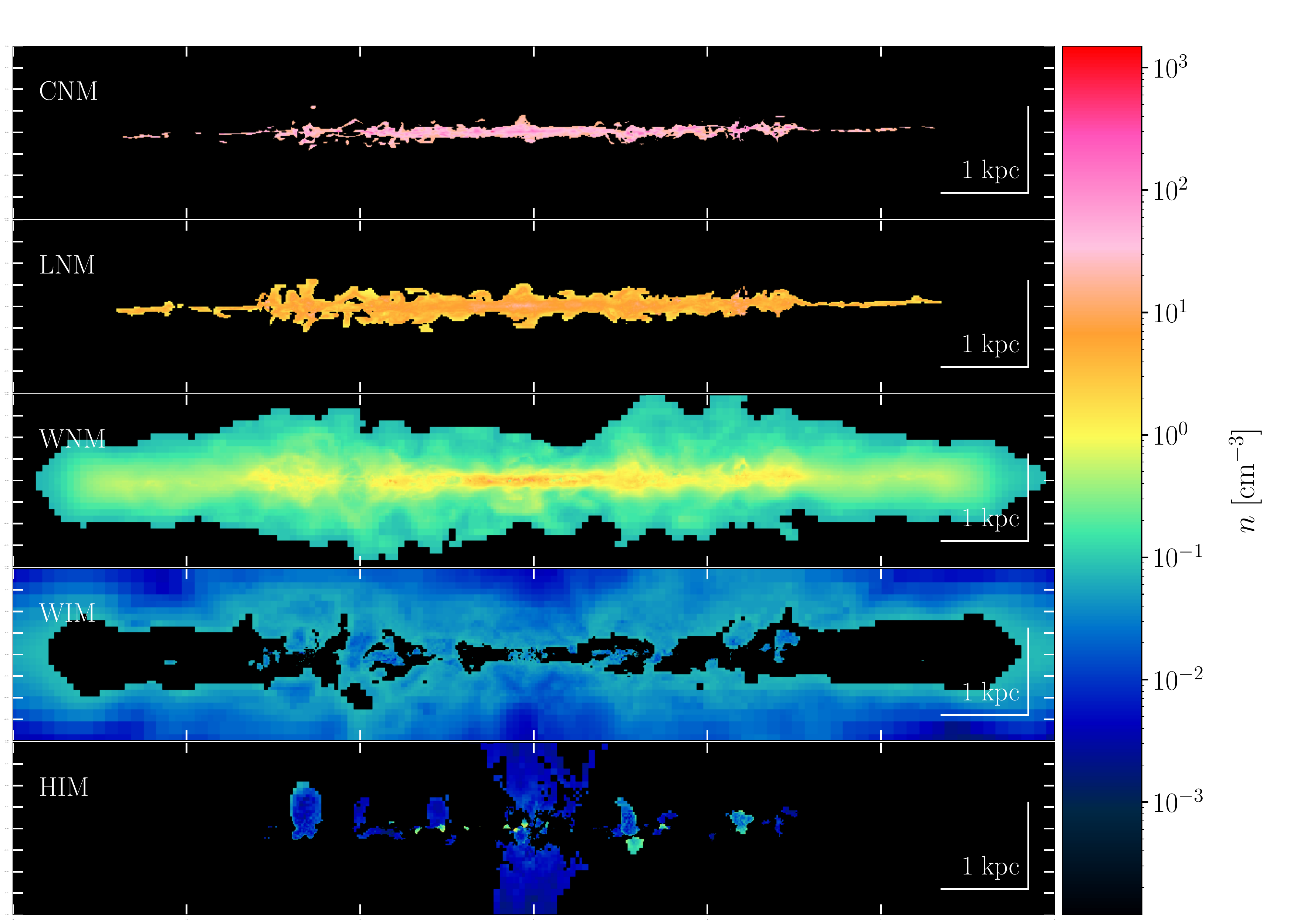}
      \vspace{1\baselineskip}
      \caption{Edge-on view of the mean gas density in a 2-kpc-thick slice through the galactic centre, for the run with anisotropic CR diffusion and $\kappa = 3 \times 10^{28}$ \cmsqs. Each panel shows one ISM phase as labelled in the plot upper left corner. They all share a common density scale.}
         \label{Fig:phasemaps}
   \end{figure}
   \begin{figure}[h!]
   \centering
      \includegraphics[width=0.45\textwidth]{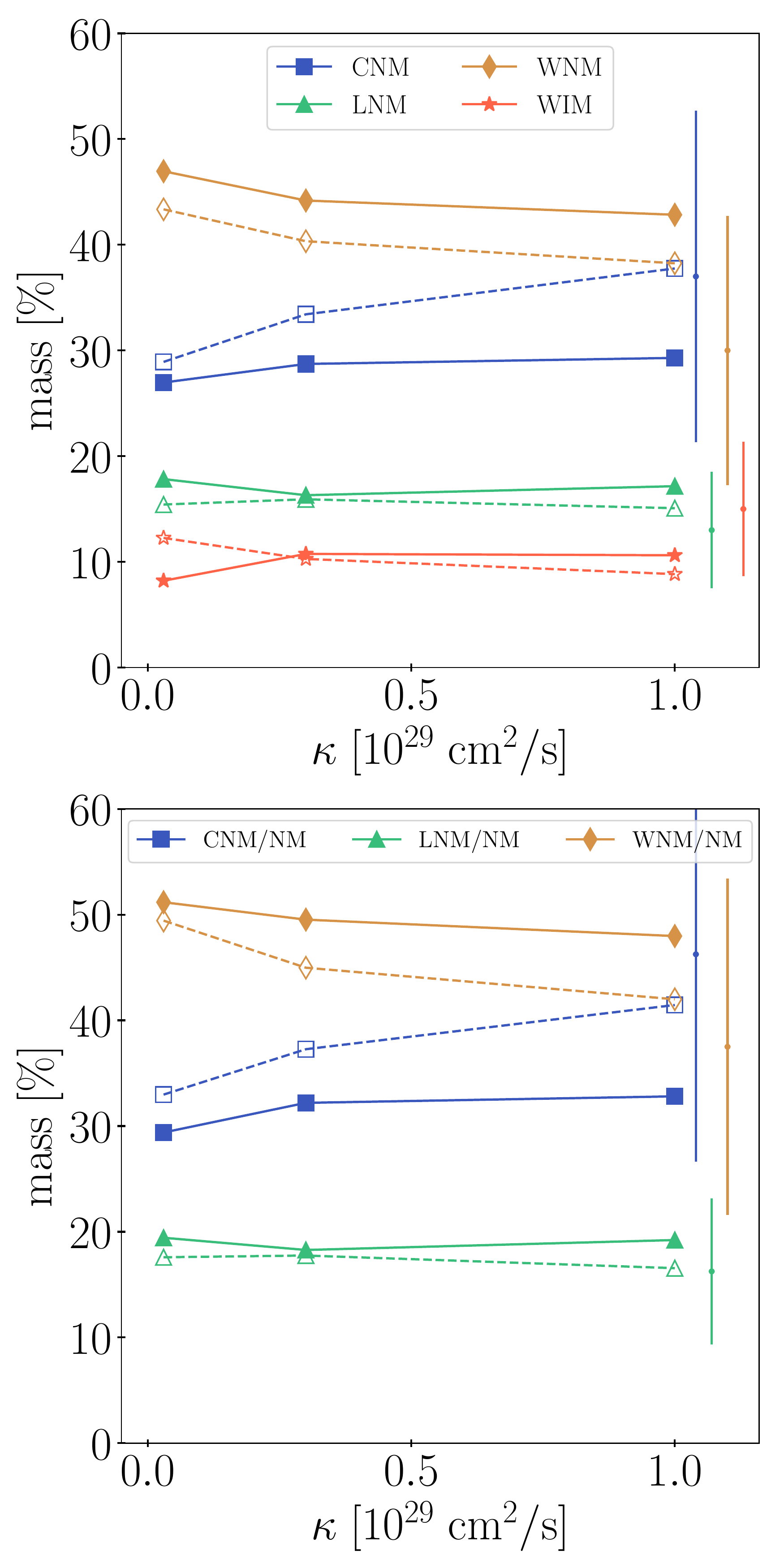}
      \vspace{-3\baselineskip}
      \vspace{2\baselineskip}
      \caption{Evolution with the diffusion coefficient, $\kappa$, of the mass fraction in different gas phases  (top) and the relative mass fractions in the neutral gas (bottom) for the isotropic (dashed) and anisotropic (solid) diffusion. The bars on the right show the ranges estimated in the Milky Way when including the atomic, dark, and molecular gas in the total neutral gas mass (NM).}
         \label{Fig:phasemassfraction}
\end{figure}
\subsection{Mass fractions in the ISM phases}\label{sec:ISMphases}
Following the multi-phase nature and nomenclature of the ISM \citep{PikelNer1968,Field1969,Mckee1977,Wolfire2003}, we divide the ISM in our simulations into five phases, three being neutral and two being ionized and much warmer than the neutral ones. The thick black lines in Fig. \ref{Fig:PhasediagPflorNoPflor} delineate their boundaries. We consider the gas to be neutral in the simulation when it has an ionization fraction lower than 20$\%$. 
The ISM phases rank with decreasing density and increasing temperature as: 
\begin{enumerate}
    \item The cold neutral medium (CNM) which gathers the densest and thermally stable gas out of which stars form. In the Miky Way, it is observed to have temperatures below a few hundred Kelvin and to account for about 30\% of the neutral \hi mass \citep{Pineda13,Murray2018,Kalberla2018}. We identify it in the phase diagram as the neutral gas (left border) with a temperature below 200 K (upper border).
    \item The lukewarm neutral medium (LNM) is the thermally unstable phase in the neutral gas \citep{Audit2005}. It accounts for about 20\% of the neutral \hi mass in the Milky Way \citep{Murray2018}. We identify it as the neutral gas (left border) with a temperature between 200~K and 5000~K.
    \item The warm neutral medium (WNM) is often the most massive gas component in a galaxy. It accounts for 50\% of the neutral \hi mass in the Milky Way \citep{Murray2018}. It gathers diffuse warm gas that we identify in the phase diagram as the neutral gas (left border) with a temperature between 5000~K and 25000~K.
    \item The warm ionized medium (WIM) corresponds to the thick disc of diffuse ionised gas exhibiting temperatures close to the WNM values and extending to large heights above the galactic disc. In the Milky Way, it corresponds to the Reynolds layer \citep{Reynolds90} with densities of 0.01-0.1 cm$^{-3}$ and a scale height of 1.8$^{+0.12}_{-0.25}$~kpc \citep{Gaensler08}. It is selected in the phase diagram as ionized gas (right border) with a temperature below 25000 K. 
    \item The hot ionized gas (HIM) corresponds to the most tenuous and hottest phase at temperatures above 25000 K in our simulations. It is produced in the simulations by the SN energy injection and takes the form of hot pockets inside the WIM. It also gathers in the galactic wind and halo, as in the Milky Way.
\end{enumerate}
In order to compare mass fractions found in the simulations and in the Milky Way, we use the following mass estimates in our Galaxy : ${\sim}8\times 10^9$~M$_\odot$ in \hi, ${\sim}10^9$~M$_\odot$ in \hd, ${\sim}10^9$~M$_\odot$ in dark neutral gas, and ${\sim}2\times 10^9$~M$_\odot$ in the WIM \citep{Ferriere01,Grenier05,Planck11, Kalberla09,Heyer15}. The uncertainties on these values are typically 30\%. As we have not simulated the transition from atomic to molecular gas, the CNM phase in the simulations includes both the atomic CNM and molecular gas of a real galaxy.


As an illustration, Fig. \ref{Fig:phasemaps} shows an edge-on view of a 2-kpc-thick central slice of each phase. Here we see how the neutral phases populate the galactic disc with vertical heights increasing with gas temperature. Part of the WNM and WIM is blown out of the disc in fountains by buoyancy, pressure gradients and momentum injection from stellar feedback. The HIM is seen in the central wind and as pockets of hot gas expanding around the SN explosion sites. 

Figure \ref{Fig:phasemassfraction} shows the evolution of the gas mass fraction present in the different phases when varying the diffusion coefficient $\kappa$ in the simulations. We also show the evolution of the neutral gas composition. The masses have been calculated in a ${\pm}4$~kpc thick disc with a radius of 6~kpc. The small mass fractions of the HIM would not be visible in this plot. They amount to 0.07\%, 0.12\%, and 0.59\% with increasing $\kappa$ value for anisotropic CR transport, and to 0.11\%, 0.22\%, 0.57\% in the isotropic case.

Despite the larger SFR per unit volume of the simulated dwarfs compared to the Milky Way, we find that the mass ranking between the phases and the mass fractions in each phase compare reasonably well with the range of values estimated in the much more massive Milky Way. So does the partition of the neutral gas between the stable and unstable phases. 

When increasing the diffusion coefficient, the WNM mass fraction is systematically reduced, but the gas moves differently to the other phases according to the degree of anisotropy in CR transport.
In the case of isotropic diffusion, the decrease in warm neutral mass represents 5\% of the total gas mass and it is primarily transferred to the CNM, adding up to 20\% more mass to the CNM and leaving the unstable LNM unchanged. Faster CR diffusion indeed leaves a deficit of CR pressure in the neutral gas which condenses more easily to the cold part of the thermal equilibrium. 
In the case of anisotropic diffusion, the magnetic field configuration reduces the loss of CR pressure at low altitude within the disc. The decrease in WNM mass fraction is slightly less important (4\%) than for isotropic diffusion, but part of the WNM gas coalesces to the CNM and part of it expands into the WIM.
The mass fraction in the unstable LNM phase is rather insensitive to changes in CR transport. 

\begin{figure}
   \centering

      \includegraphics[width=0.45\textwidth]{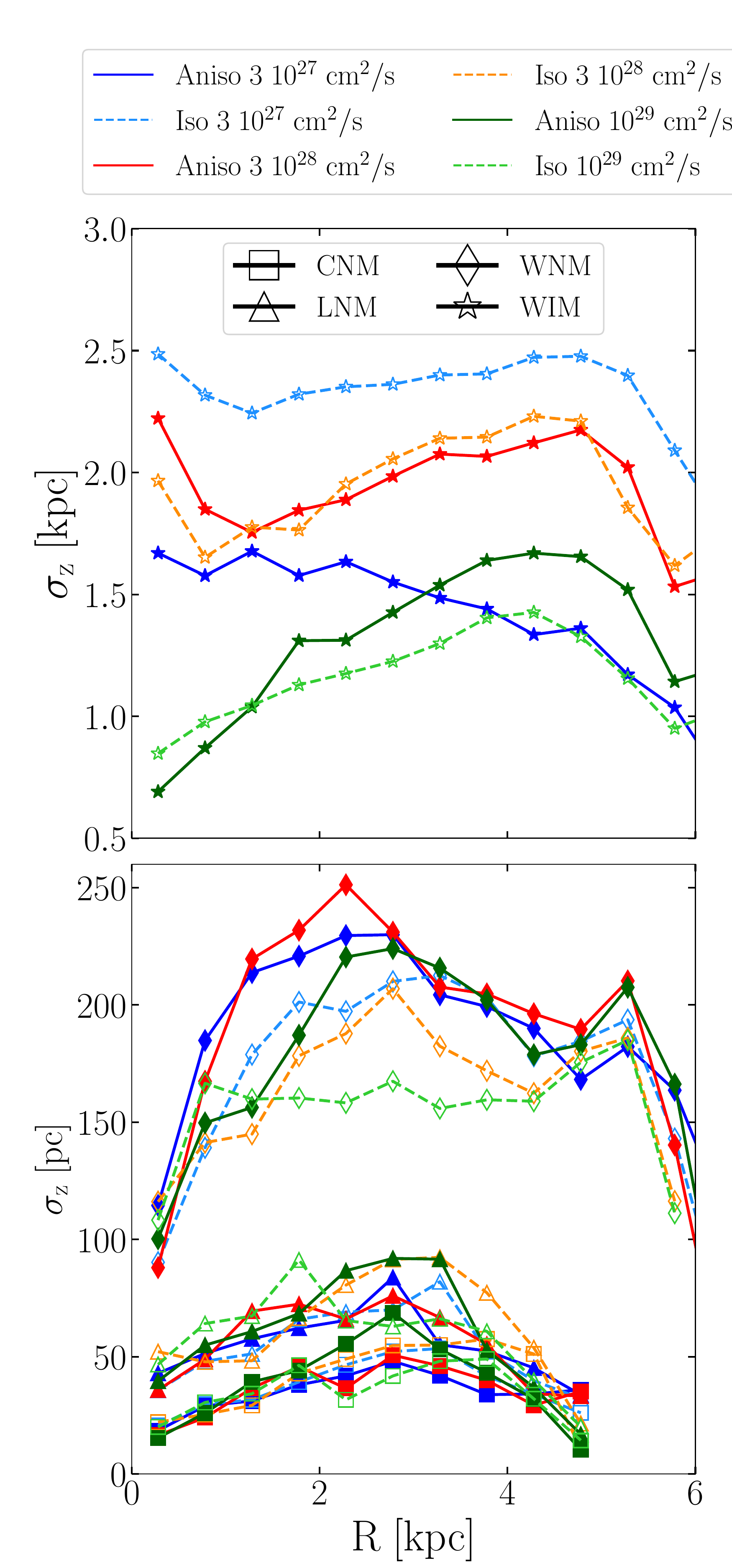}

      \vspace{1\baselineskip}
      \caption{Vertical scale height of the gas mass in the different phases. The upper panel is in kpc instead of pc due to the larger scale of the WIM. The colours code the different CR transports.}
         \label{Fig:phasescaleheight}
   \end{figure}
   
 \begin{figure*}[ht!]
   \centering

      \includegraphics[width=0.95\textwidth]{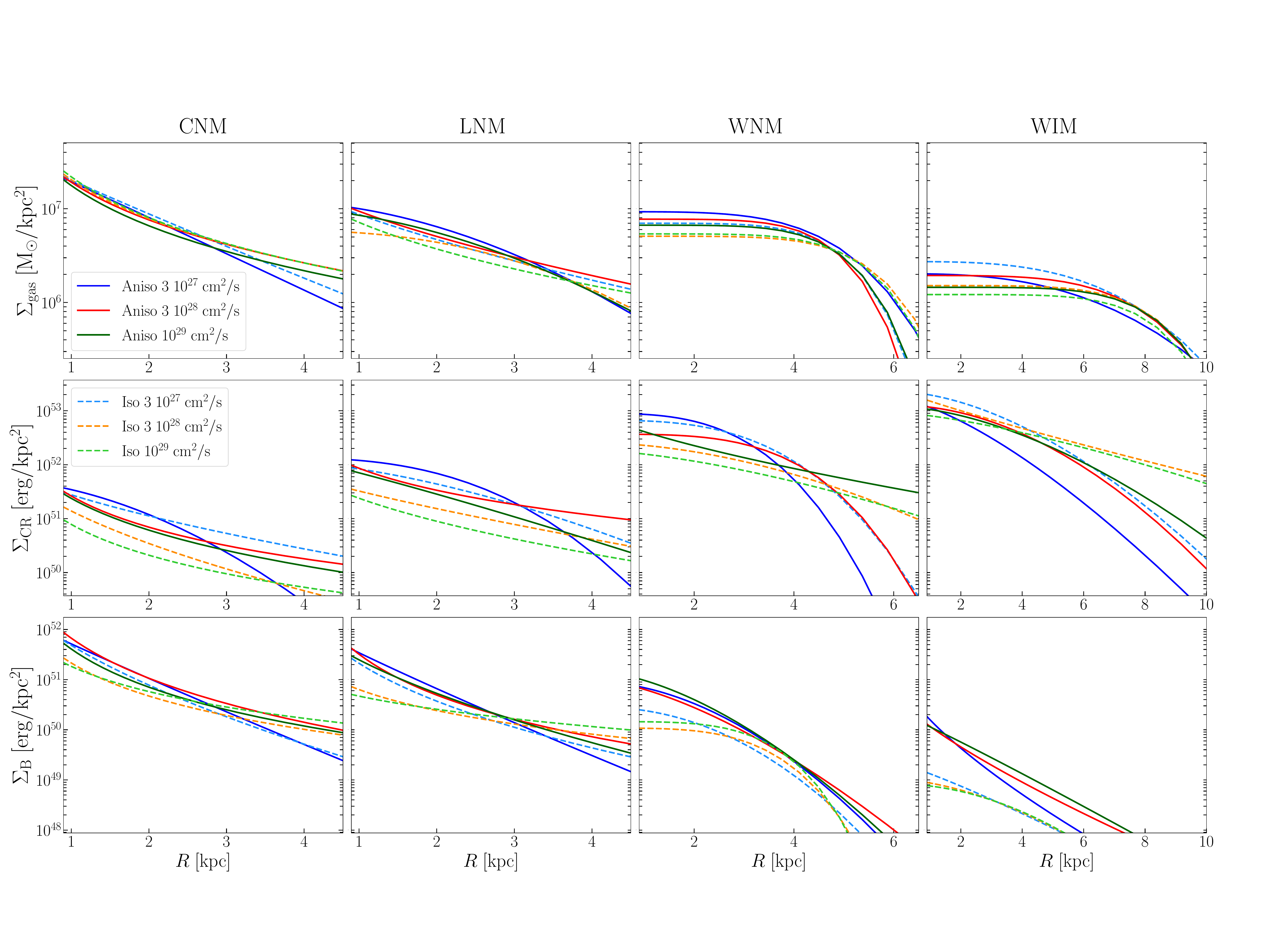}
      \vspace{-3\baselineskip}
      \vspace{2\baselineskip}
      \caption{S\'ersic fits to the surface density profiles in gas mass (top), CR energy (middle), and magnetic energy (bottom) as a function of radial distance in the disc. The fits were obtained for the different ISM phases labelled above the columns. 
      }
         \label{Fig:phaseradialprofiles}
\end{figure*}

\subsection{Large-scale gradients in the ISM phases}\label{sec:gradphase}
In order to look at the vertical extent of the ISM phases, we have calculated the vertical scale heights of the gas mass in each phase according to eq. \ref{eq:verticaldisp}, in axisymmetric radial bins. The results are shown in Fig. \ref{Fig:phasescaleheight}. For standard diffusion with $\kappa = 3 \times 10^{28}$~cm$^2$~s$^{-1}$, the WIM and WNM heights compare favourably with the Milky Way values of 1.8$^{+0.12}_{-0.25}$~kpc for the WIM \citep{Gaensler08} and of 150-300~pc for the WNM in the non-flaring regions of our Galaxy \citep{Kalberla09}. The denser phases are, however, about twice more condensed about the midplane than in our galaxy if we compare to the $\pm$100-pc thickness of the molecular layer in the Milky Way \citep{Heyer15}. This flattening occurs despite the larger stellar feedback per unit volume in the simulated dwarfs compared to the Milky Way.

All gas phases significantly respond in height to changes in CR transport, by 30\%-40\% for the neutral phases, and by as much as a factor of 2 for the more diffuse and extended WIM, but we find no systematic trend between the isotropic and anisotropic cases, nor with increasing diffusion coefficient, even in the most massive WNM phase.  Different regions respond differently, depending on the location in the galaxy. This diversity likely results from local variations in the late history of SN feedback and CR injection. We will investigate the variability of these scale heights in the companion paper.

In order to study the radial evolution of the gas mass and of the CR and magnetic energies in  different ISM phases, we have fitted S\'ersic profiles to the surface density of each of those quantities in concentric annuli, within $\pm$3-kpc from the midplane, for radii larger than 1 kpc. The vertical thickness of the annuli is not important as long as it includes the phases we are interested in (up to the WIM). The resulting fits are presented in Fig. \ref{Fig:phaseradialprofiles}. We note that the mean surface density in gas mass and in magnetic energy decreases from the dense to the diffuse phases despite the broader scale heights because the gas density and magnetic-field strength strongly drop in the diffuse regions. The opposite occurs with the CR-energy surface density because the CR scale height exceeds that of the gas, even for the thick WIM. 

We find that the mass surface density in the three neutral phases is rather stable, in magnitude and in radial profile, against changes in CR transport. An overall systematic decrease can be seen in gas and magnetic surface densities in the WNM between the anisotropic and isotropic cases. It relates to the corresponding decrease in gas scale height that is visible in Fig. \ref{Fig:phasescaleheight}.
The CR energy surface density drops more markedly at the periphery of the galaxy in the diffuse WNM and WIM phases when CRs diffuse along the magnetic field because the magnetic gradients across the galaxy are steeper in that case (see Fig. \ref{Fig:EnergyMapsFaceon}).

\section{Results on star formation}
\label{sec:SFR}
We present the star-formation activity in the six galaxy realisations, first focusing on the global SFR in the galaxy and the Kennicutt-Schmidt (KS) relation within the galaxy, then discussing how CR transport affects star formation.

\begin{figure}
   \centering
      \includegraphics[width=0.45\textwidth]{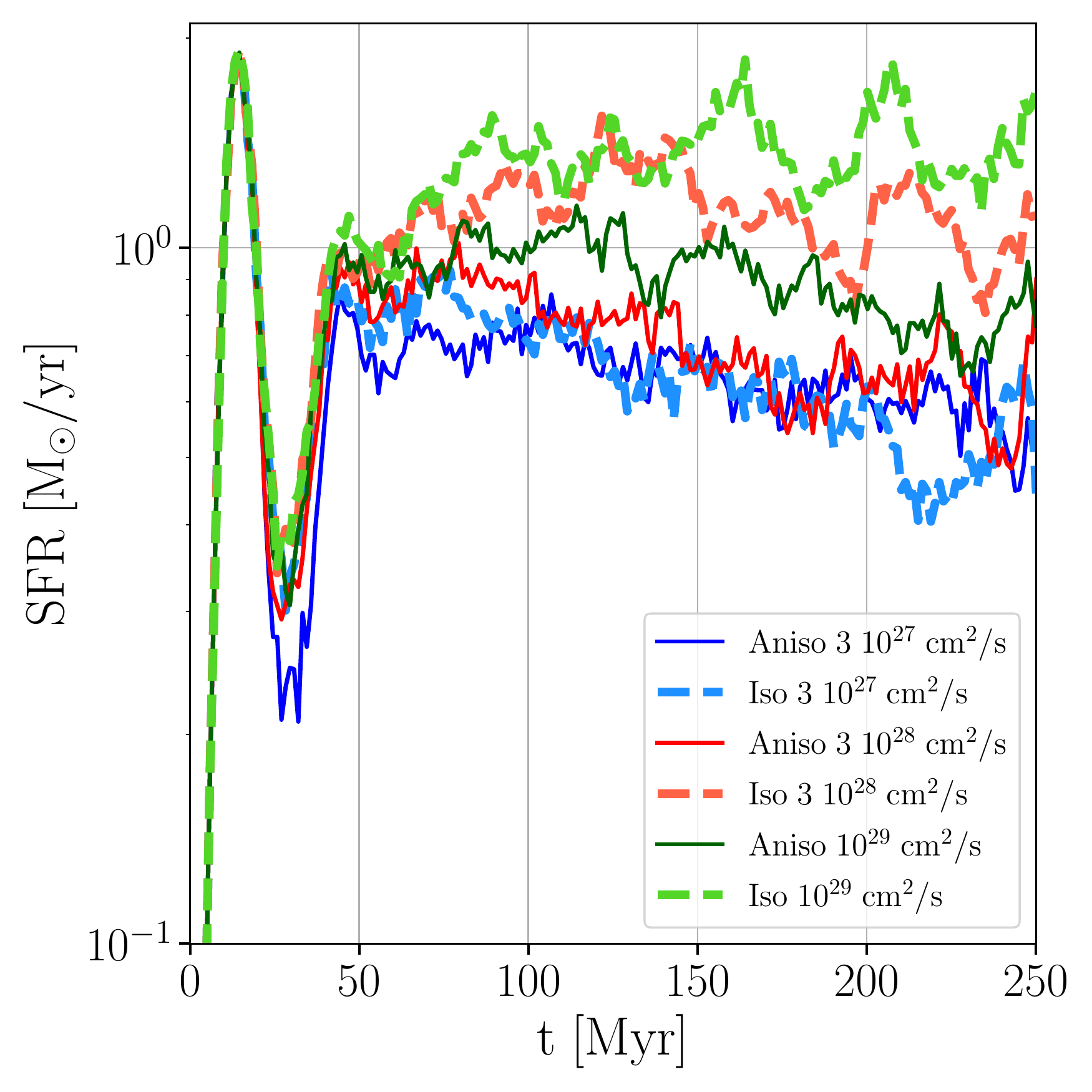}
      \vspace{-3\baselineskip}
      \vspace{2\baselineskip}
      \caption{Star-formation rate for all the runs up to 250 Myr.}
         \label{Fig:SFR}
   \end{figure}
\begin{figure}
   \centering
      \includegraphics[width=0.45\textwidth]{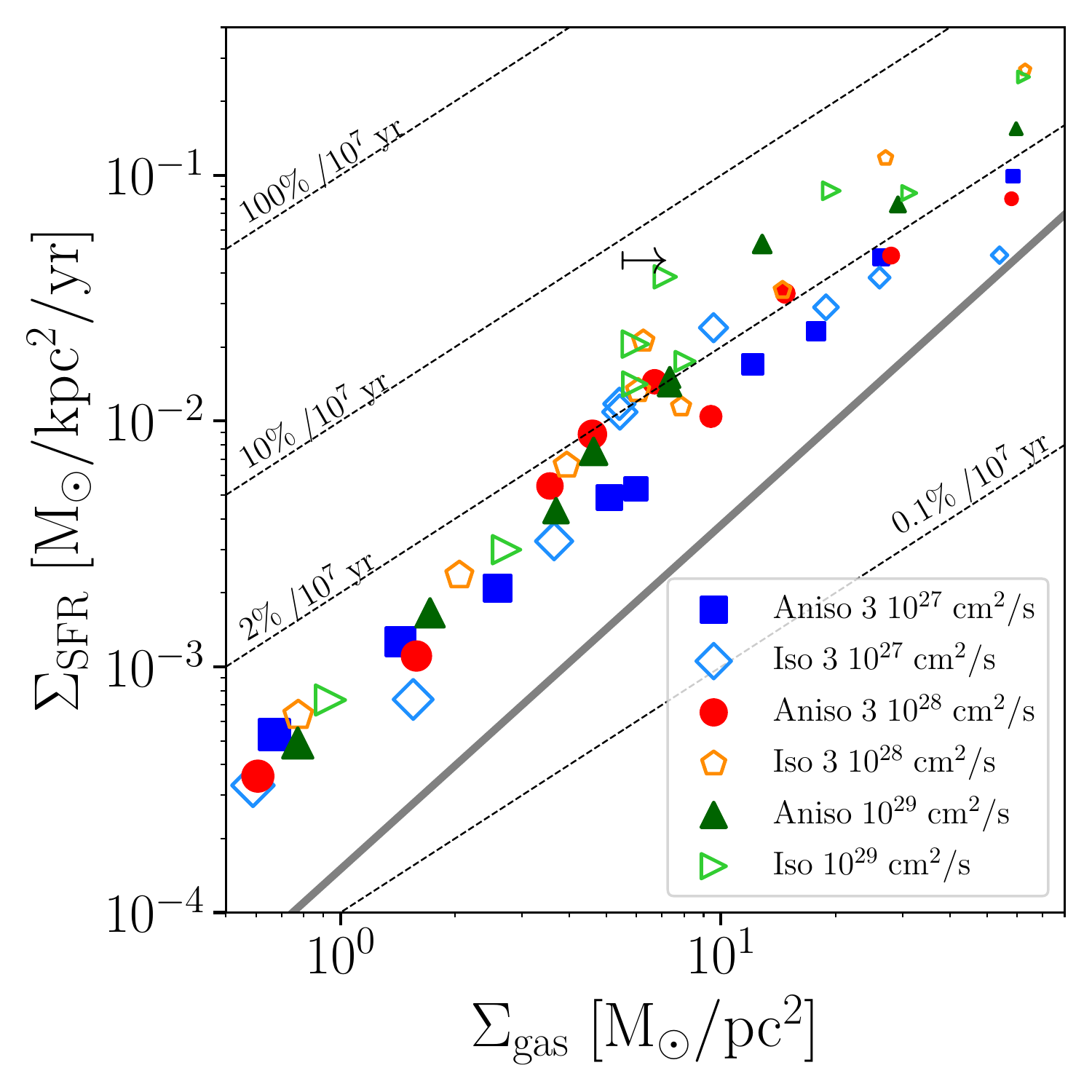}
      \vspace{-3\baselineskip}
      \vspace{2\baselineskip}
      \caption{Relationship between the surface densities of star formation and gas mass in the CNM phase, in galactocentric rings of 500~pc in radial width. The solid grey line shows the Kennicutt-Schmidt relation \citep{Kennicutt1998}. To show the evolution of this relation inside the galaxy, the marker size scales with the distance to the centre of the galaxy. Additionally, the black arrow delimits the lower gas mass surface density for annuli inside 2~kpc. The dotted lines show different values of constant SFR in units of percent per 10 Myr.}
         \label{Fig:KSlaw}
   \end{figure}

\subsection{Star formation rate and Kennicutt-Schmidt relation}
\label{sec:SFRandKS}

Figure \ref{Fig:SFR} presents the time evolution of the global SFR for all the runs. It shows that, as CR diffuse out faster, both the reduced CR pressures in the CNM (discussed below) and the increased gas mass fraction present in the CNM (Fig. \ref{Fig:phasemassfraction}) concur to increase the SFR. The SFR response to CR transport is highly non linear. It is consistently around ${\sim}0.6$~M$\odot$/yr if the CRs remain concentrated around their birth place in the CNM, either because of slow diffusion ($3\times 10^{27}$~cm$^2$/s, isotropic or not) or for ten times faster diffusion as long as the magnetic field prevents the CR flux from rapidly diverging (anisotropic $3\times 10^{28}$~cm$^2$/s). The SFR is increased by a factor up to 2.5 for fast isotropic diffusion ($\kappa \geq 3\times 10^{28}$~cm$^2$/s). 

SFR surface densities, $\Sigma_{\textrm{SFR}}$, are observed to scale as a power law of the total gas-mass surface density, $\Sigma_{\textrm{gas}}$, with an index of 1.4 when averaged over the disks of many types of galaxies: $\Sigma_{\textrm{SFR}}\propto \Sigma_{\textrm{gas}}^{1.4}$  \citep{Kennicutt1998,Kennicutt12}. Figure \ref{Fig:KSlaw} shows the relation we find in the different simulations between the SFR and gas surface densities calculated in concentric rings paving the galaxies to an outer radius of 6~kpc. The rings are 500~pc in radial width and include only the CNM gas. The symbol sizes in the plot increase with radial distance in the disc. The black arrow marks the lowest gas-mass surface density found in the inner rings, within  R$<2$~kpc of the galaxy centre. Symbols to the right of this arrow therefore describe star formation in the inner regions, inside the co-rotation radius. 
The dashed lines correspond to a constant SFR per unit gas mass and per time interval of 10 Myr. The choice of 10 Myr is arbitrary, but it corresponds roughly to the free-fall time for gas in the CNM, more specifically, gas at a temperature below 200 K and densities above $\sim10\,\rm cm^{-3}$. 

The star formation strategy in our simulations being based on the Kennicutt-Schmidt (KS) relation, we find SFR surface densities that globally follow the KS trend. They are four times larger than the formal KS relation, but close to the constant $\epsilon_{*} = 2\%$ efficiency used per star-forming cell in the simulations (see section \ref{sec:SimMethods}), in particular in the inner regions ($R<2$~kpc, to the right of the black arrow in the plot). We note that CR transport does not influence star formation beyond about 2~kpc whereas star formation is suppressed for anisotropic or slow diffusion in the inner regions. We explore the origin of this dichotomy in the following section.   

\subsection{CR feedback on star formation}\label{sec:CRandSFR}
\begin{figure*}
   \centering
      \includegraphics[width=0.9\textwidth]{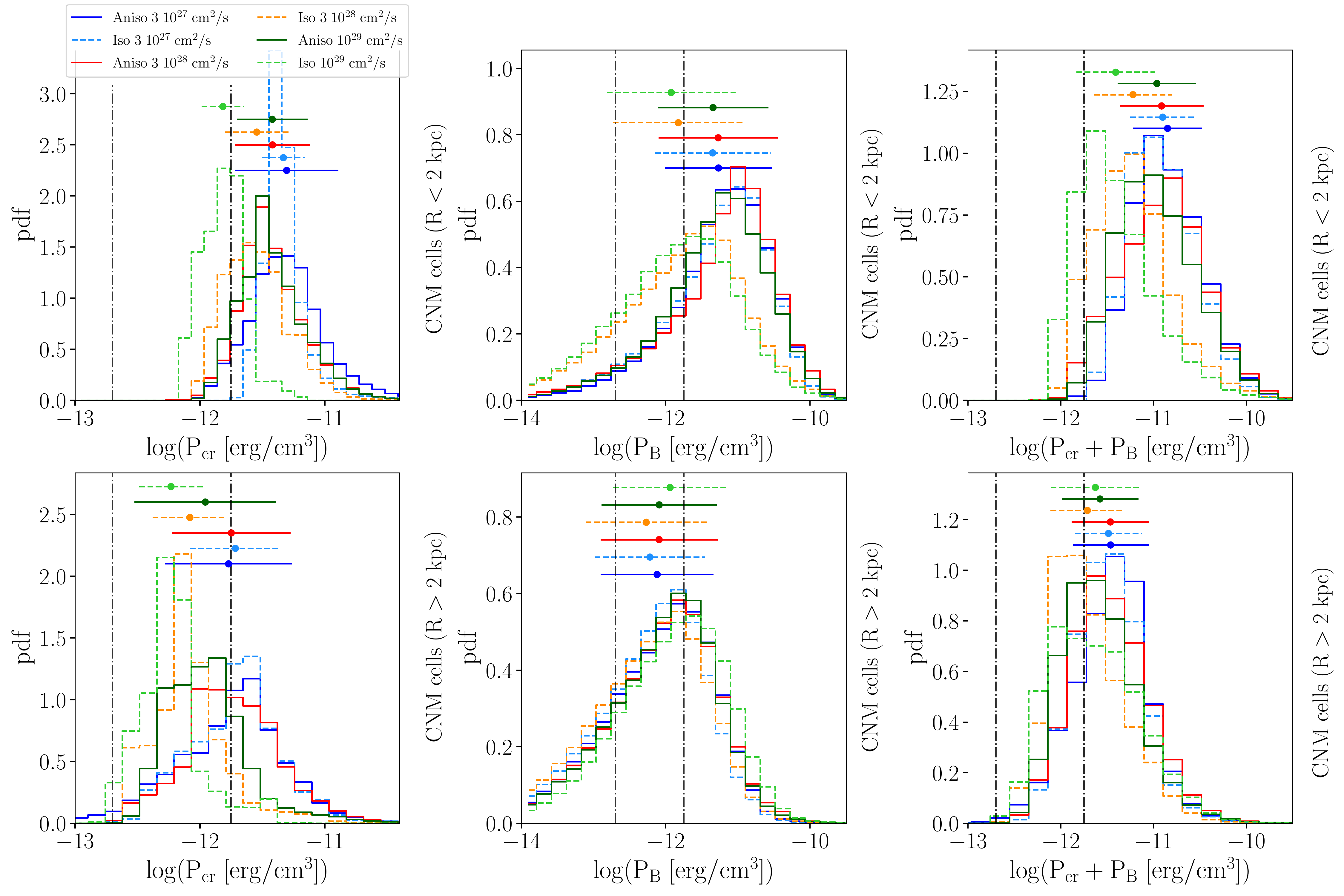}
      \vspace{-3\baselineskip}
      \vspace{2\baselineskip}
      \caption{Normalized distribution of the number of CNM cells for CR (left) and magnetic (center) pressures and the sum of both pressures (right), separated in two regions: inside 2 kpc (top) and outside 2 kpc (bottom). The error bars show the mean and dispersion of each distribution. The vertical dash-dotted lines indicate the range of total pressure where the thermal instability can supply the CNM. }
         \label{Fig:Pcrdist}
\end{figure*}

As mentioned above, a cell must be denser than 80 atoms per cm$^3$ and colder than 3000~K to be considered for star formation. This will occur almost exclusively in cells that belong to the CNM. CR transport affects the mass inflow to star-forming CNM cells in different ways. They act on dense cloud scales by adding pressure directly or indirectly via magnetic amplification, to hold the clouds against gravity. Their pressure gradients also act at the scale of cloud complexes to regulate the mass flux from gravitationally unbound to bound entities. When going from fast to slow CR diffusion, the observed decrease in CNM mass fraction helps suppress star formation, but the less than 40\% change in mass fraction is not sufficient to explain a factor up to 2.5 drop in SFR. 
As we have noted in section \ref{sec:gradients} that the magnetic-field strength distribution significantly responds to CR transport, both radially along the disc and in strongly magnetised filaments in the inner galaxy, we further explore the dual impact of CR and magnetic pressure on SFR. 

Figure \ref{Fig:Pcrdist} shows the normalized distributions of CR, magnetic pressures  and the combinations of both, found in CNM cells located in the inner disc (top row) and in the outer disc (bottom row). We also plot an error bar giving the mean value and rms dispersion of each distribution. We find a systematically larger spread in CR pressures in the anisotropic diffusion runs compared to the isotropic cases because the spatial distribution of CRs is more contrasted (see Fig. \ref{Fig:EnergyMapsFaceon}). Conversely, the overall range of magnetic pressures in the CNM appears to be rather insensitive to CR transport. 

Figure \ref{Fig:Pcrdist} shows a marked shift toward lower CNM pressures, both in CR and in magnetic pressures, for faster diffusion and in particular for fast isotropic transport. The mean CR pressures in both the outer and inner disc vary by a factor up to 2.6. The mean magnetic pressures in the inner disc are 3.5 times lower for the isotropic $\kappa \geq 3\times 10^{28}$~\cmsqs cases than for the others. 
The change in CNM CR pressure is due to faster particle propagation out of the clumpy CNM where they are produced. They diffuse more quickly to the more extended WNM and WIM phases. The origin of the magnetic response is less clear. The face-on maps in Fig. \ref{Fig:EnergyMapsFaceon} exhibits differences by an order of magnitude in magnetic energy density in the inner galaxy structures. \cite{Pakmor16} had earlier noted that the timescale for magnetic amplification in the disc could change by a factor of ten between isotropic and anisotropic CR transport for the same diffusion coefficient ($10^{28}$~\cmsqs). 
In the model of \cite{Shukurov06} for the evolution of the toroidal field, 
vertical fountain flows can efficiently remove magnetic helicity in the disc to allow the rapid growth of the azimuthal field. Following this model, \cite{Pakmor16} proposed that the amplification rate relates to the strength of the velocity and magnetic gradients perpendicular to the plane, shallower gradients inhibiting magnetic-field amplification. Figure \ref{Fig:scaledistances} shows, however, that the magnetic scale height is roughly twice lower for isotropic diffusion than for the corresponding anisotropic case. On the other hand, Fig. \ref{Fig:EnergyMapsEdgeon} shows that more magnetised gas is pushed off the plane, particularly above the inner regions, in the cases of anisotropic or very slow CR transport. Given this complex picture, we defer a detailed study of the CR feedback on the magnetic-field distribution to future work.

Reduced pressures in CRs and in the magnetic field both concur to provide less support to the clouds against  gravitational collapse to reach the dense star-forming phase. 
Figure \ref{Fig:Pcrdist} indicates that the pressure distributions in the CNM are indeed rather similar for the three cases exhibiting the lowest SFR curves. In most of the cells, the CR or magnetic pressures alone can keep the gas well above the pressure domain where the thermal instability between the WNM and CNM is fully operational to build dense clouds. This range corresponds to $1500 \lesssim P \lesssim 12500$~K~cm$^{-3}$ in Fig. \ref{Fig:PhasediagPflorNoPflor} and it is marked as vertical lines in Fig. \ref{Fig:Pcrdist}.  
For the three "low-SFR" CR transport cases, the CR pressures remain large enough in the vicinity of CR sources to prevent further gas coalescence and the propagation of star formation.    
For the two cases with fastest CR dilution, the global SFR almost linearly responds to the number of cells that can take advantage of the thermal instability to transfer gas from the WNM to the CNM phase.

The fact that the change in CR and magnetic pressure distributions is more pronounced in the inner disc, within a radius of ${\sim}2$~kpc, corresponds to the larger changes seen in efficiencies in the KS plot for the inner rings compared to the outer ones in Fig. \ref{Fig:KSlaw}. \cite{Hanasz09} showed that random magnetic perturbations injected by SN explosions can be efficiently ordered and that they can grow to large-scale fields up to the equipartition level with an e-folding timescale that is close to the disk rotation period. We note, however, in Fig. \ref{Fig:Pratios} that, in the cases of fastest CR dilution where the SFR (hence the SN rate) is highest, the magnetic pressure profile saturates at a level that is two to three times below that obtained for slower or more confined CR transport. A low field is reached even though the galaxies have experienced large SFRs over nearly a rotation period, i.e. an e-folding time in growth.

%

Figure \ref{Fig:SFR} also shows that temporal variations in the SFR have larger amplitudes in the three cases where the CRs rapidly spread away from their sources. The cold gas appears to be concentrated in fewer, but more fully developed spiral arms in the cases of lower CR pressure resistance (see Fig. \ref{Fig:EnergyMapsFaceon}). 
In the companion article we will further explore the impact of CR pressure on the gas circulation between the WNM reservoir and CNM clouds, and if the degree of CR clustering acts at a preferential spatial scale to restrain star formation.

Another interesting effect is the decline in SFR except for the two runs with fast, isotropic CR diffusion ($\kappa \geq 3 \times 10^{28}$~\cmsqs) for which the average SFR on long time scales remains constant after 100 Myr. Since the total amount of gas in the simulation volume is virtually unchanged amongst the different runs, thus ruling out the possibility that this decline is due to gas escape from the box, we speculate the cosmic ray pressure affects the diffuse phases of the ISM, not allowing them to condensate into denser states. Alternatively, we consider the possibility that this decline reflects a longer timescale to reach a steady state in simulations with slow diffusion, however, we still observe this trend in the SFR 50 Myr after the present analysis.

\section{Results on $\gamma$-ray luminosities}
\label{sec:gamma}

\begin{figure}
      \includegraphics[width=0.95\linewidth]{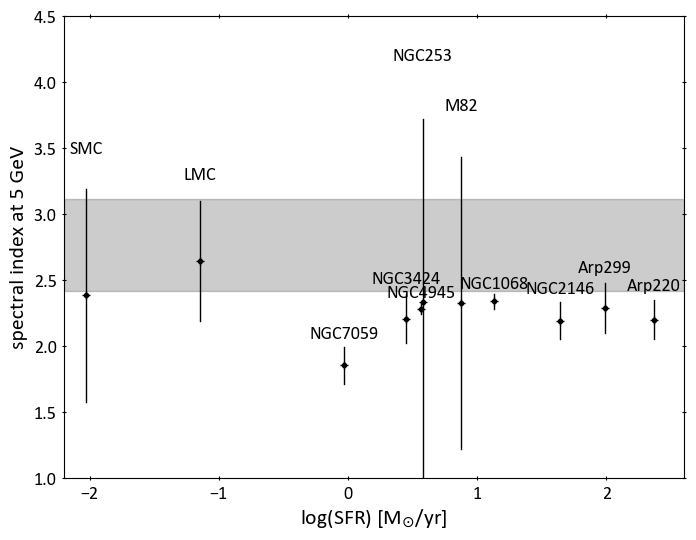}
      \vspace{2\baselineskip}
      \caption{Spectral evolution of the \g-ray emission as a function of SFR in the current sample of galaxies detected by Fermi LAT. The spectral indices are estimated at an energy of 5~GeV to take advantage of the good telescope sensitivity to point sources at this energy as well as the potential dominance of the hadronic contribution to the total emission in this range (reduced pulsar and CR electron contributions). The grey band marks the 68\% confidence range for the same spectral index in the hadronic emission observed around the Sun and in other regions of the Milky Way.}
         \label{Fig:specgam}
\end{figure}

\begin{figure*}
   \centering
      \includegraphics[width=0.95\textwidth]{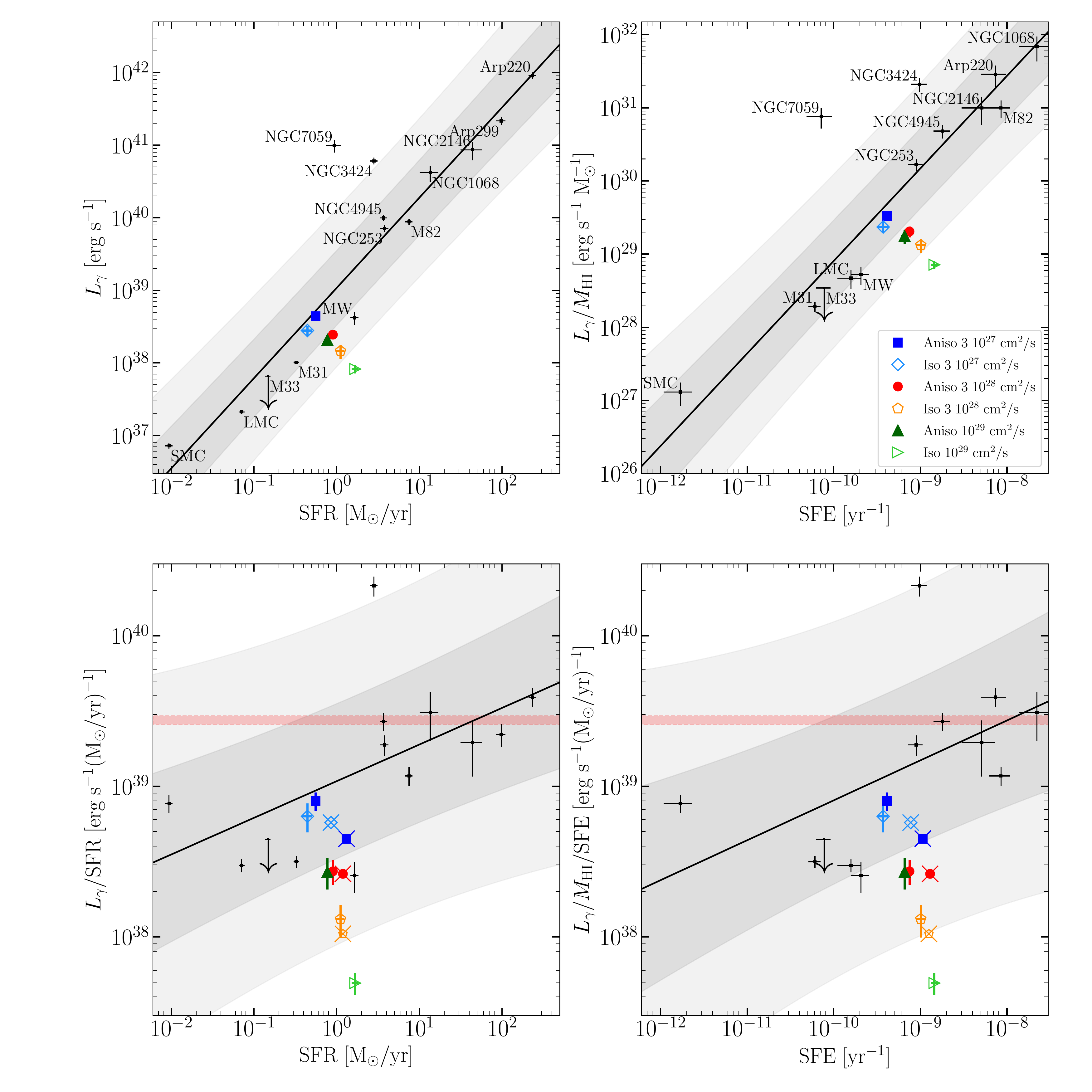}
      \vspace{-3\baselineskip}
      \vspace{2\baselineskip}
      \caption{Evolution of the \g-ray luminosities, $L_{\gamma}$, in the 0.5-50 GeV band as a function of the galaxy SFR (top left) and the same evolution per unit gas mass and star formation efficiency, SFE=SFR/$M_{\rm H}$ (top right). The lower panels show the same trends divided by the SFR and SFE, respectively. The calorimetric limit for CRs with energies between 5 and 500 GeV is shown as the shaded red band. The black data points show Fermi LAT detections (and upper limit as an arrow for M33). NGC 7059 lies beyond the upper bound of the lower plots. The coloured points give the simulation results with error bars reflecting the rms dispersion in \g-ray luminosity and in SFR due to time variations over 150~Myr. The results of the simulations obtained by \cite{Dashyan20} for the same CR assumptions, but warmer monophasic gas, are added to the lower panels and crossed out with an X. The best fits to the Fermi data (ignoring NGC 7059) are shown as black solid lines, with their 68\% (dark grey) and 95\% (light grey) confidence regions, respectively.}
         \label{Fig:gammaLum}
   \end{figure*}

CR nuclei with kinetic energies above 1 GeV per nucleon predominantly loose their energy in hadronic interactions with gas nuclei as they diffuse through the ISM. The neutral pions produced in such interactions decay into \g rays. We have calculated the global \g-ray luminosities of the simulated galaxies in the 0.5-50~GeV band to reduce the bremsstrahlung and inverse-Compton contributions from unmodelled CR electrons to a negligible level. We have assumed the spectrum of the different CR nuclei to be the same as recently inferred in the solar neighbourhood from AMS-02 and Voyager data \citep{Boschini20}. The spectra of the different nuclei, with atomic numbers $Z \leq 28$, jointly yield an energy density $e_{CR\odot} = 1.4\times 10^{-12}$~erg~cm$^{-3}$ for kinetic energies above 1 GeV/nucleon. 

Uncertainties of order 30\% currently exist in the cross sections for \g-ray production in hadronic interactions \citep[see Fig. 3a of][]{Grenier22}, so we have adopted the average emissivity spectrum, $q_{\gamma}(E_{\gamma})$ from \cite{Casandjian15}, that has been directly measured per gas nucleon in the local atomic gas with the Large Area Telescope on board the Fermi Gamma-Ray Space Observatory (aka Fermi LAT). This spectrum is consistent with the CR spectra reaching the heliosphere and it yields a total \g-ray luminosity per gas nucleon in the 0.5-50 GeV range : 
\begin{equation}
    L_{\gamma H} = \int_{0.5\,\textrm{GeV}}^{50\,\textrm{GeV}} 4\pi E_{\gamma} q_{\gamma}(E_{\gamma}) dE_{\gamma} = 1.1\times 10^{-28}~\textrm{erg~s}^{-1}~\textrm{H-atom}^{-1}
\end{equation}
\noindent This luminosity takes the energy-dependent composition of the CRs into account, but one can expect CR spectral variations across a galaxy due to the energy dependence of the diffusion coefficient, $\kappa \propto E_{CR}^{\delta}$ and the proximity of CR sources. Estimates of the $\delta$ index range between 0.3 and 0.6 in the Milky Way \citep{Gabici19}, so one would expect marked spectral gradients around CR sources, but \g-ray observations indicate only a small progressive softening of the CRs (by ${\sim}0.2$ in spectral index) from the inner molecular ring to the outskirts of the Milky Way \citep{Acero16diff,Pothast18}, and it is unclear whether the 0.1 hardening seen toward the inner regions is due to point-source confusion or to harder CRs diffusing near star-forming regions. The same emissivity spectrum as the local one, with only small variations in amplitude and no appreciable changes in spectrum, has been found in a variety of Milky Way environments, in particular in and out of star-forming spiral arms \citep[see Fig. 4 of][]{Grenier15} and up to 1~kpc in height above the Galactic plane \citep{Tibaldo15}. No spectral changes have either been found outside active CR acceleration sites such as the Cygnus-X superbubble \citep{Ackermann12Cyg}.

Furthermore, Fermi LAT has detected the global \g-ray emission from a sample of galaxies spanning four orders of magnitude in SFR. Figure \ref{Fig:specgam} shows that their emission spectrum is consistent with the interstellar emission of the Milky Way at low SFR, but it is systematically harder at large SFR. The spectral index shows, however, no sign of evolution with the SFR, hence with the CR production rate, even if the galaxy is ten to a hundred times more active than the Milky Way. This lack of evolution, together with the apparent spectral uniformity of the hadronic emission in many places of the Milky Way, and the fact that our simulated galaxies exhibit moderate SFRs close to the Milky Way rate, prompted us to use the local-ISM \g-ray emissivity spectrum, weighted by the ambient CR energy density, throughout the simulated galaxies.

Under this assumption, the total hadronic \g-ray luminosity over the volume $V$ of a galaxy can be expressed as : 
\begin{equation}
\label{eq:Lgam}
\begin{split}
    L_{\gamma} & = \int_V n_H(\mathbf{r}) \frac{e_{CR}(\mathbf{r})}{e_{CR\odot}} dV \int_{0.5\,\textrm{GeV}}^{50\,\textrm{GeV}} 4\pi E_{\gamma} q_{\gamma}(E_{\gamma}) dE_{\gamma} \\
  & =  \frac{L_{\gamma H}}{e_{CR\odot}} \int_V n_H(\mathbf{r}) e_{CR}(\mathbf{r}) dV
\end{split}
\end{equation}
\noindent where $n_H$ notes the number density of gas atoms and $e_{CR}$ is the CR energy density. 

\subsection{The \g-ray to SFR relation}
\label{sec:gammaSFR}

Figure \ref{Fig:gammaLum} presents how the global \g-ray luminosities of the simulated galaxies 
relate to their SFR and how those galaxies compare with the sample observed with Fermi LAT. The derivation of the different stellar and gas properties used in Fig. \ref{Fig:gammaLum} are detailed in appendix \ref{sec:galprop} and the values are listed in Table \ref{tab:gammaSFR}. Compared to previous publications about the \g-ray-FIR trend, we have updated the Fermi LAT data points using the latest source information obtained for the DR3 version of the 4FGL catalogue \citep{Abdollahi224FGLDR3}. We stress that the detection of M33, proposed by \cite{Ajello20} and used in subsequent studies \citep[e.g.][]{Kornecki2020,Werhahn21}, has not been confirmed in 4FGL-DR3 despite the improved sensitivity provided by longer observations (twelve years of LAT data instead of ten).

Because the CR production rate scales with the SN rate, hence with the SFR, we expect a strong correlation between the \g-ray luminosity and the SFR \citep{Ackermann2012}. This correlation is prominent in the upper left panel of Fig. \ref{Fig:gammaLum}. It corresponds to $L_{0.5-5\,GeV} = 10^{39.03 \pm 0.14}  ({\rm SFR}/[M_{\odot}/yr])^{1.24 \pm 0.11}$~erg/s. We detail this trend by showing how the ratio of the \g-ray luminosity over the SFR evolves with the latter in the bottom left panel. The SFR relates to the gas mass surface density in the disc, but the \g-ray luminosity scales with the gas mass through which CRs diffuse, which encompasses a much broader volume than where the stars form. We thus present in the right panels of Fig. \ref{Fig:gammaLum} how the \g-ray luminosity per gas mass evolves with the star-formation rate per gas mass, i.e. the star formation efficiency, SFE. The best-fit corresponds to $L_{0.5-5\,GeV}/M_{\rm HI} = 10^{41.6 \pm 1.2} ({\rm SFE}/yr^{-1})^{1.27 \pm 0.13}$~erg~s$^{-1}$~M$_{\odot}^{-1}$. 
Since we have only considered atomic gas mass estimates for the real galaxies to ensure a better uniformity in the sample (as discussed in section \ref{sec:galprop}), the SFE presented in Fig. \ref{Fig:gammaLum} should be considered as upper limits, of order 30\%-50\% above the actual values. The strong correlation seen between the Fermi LAT luminosities and the SFEs in the upper right panel of Fig. \ref{Fig:gammaLum} is driven by the SFR dependence of the total CR power produced in a galaxy. Power-law fits to the different trends have been performed and we show the 95\% confidence region around them. One galaxy, NGC 7059, stands at variance in Fig. \ref{Fig:gammaLum}. Its high luminosity-to-SFR ratio and its particularly hard spectrum cast doubts on its association with the 4FGL2127.6-5959 LAT source even though the spiral galaxy is located at the edge of the 68\% error region of the LAT source, in an uncrowded region at high latitude. 

Figure \ref{Fig:gammaLum} also presents the calorimetric limit we expect from our choice of CR nuclei spectra and \g-ray emissivity. We detail its derivation as it provides insight into the main parameters that influence the total \g-ray luminosity\footnote{We follow the derivation presented in section 5.1 of \cite{Ackermann2012}}. A galaxy becomes a good “calorimeter” for CR nuclei when their diffusive residence time in the ISM, $\tau_{res}$, exceeds their loss timescale due to hadronic interactions with interstellar gas, $\tau_{pp}$. When so, the CR energy is reprocessed inside the galaxy. 
For a CR with kinetic energy $E_k$ and velocity $\beta(E_k) c$, the hadronic interaction timescale varies with the ambient gas density as $\tau_{pp} n_H =  [\beta(E_k) \,c \,\sigma_{inel}(E_k)]^{-1}$. The \g-ray luminosity of the Milky Way peaks
in the 0.1-10 GeV band \citep{Strong2010} and the
photons are typically produced by ten times more energetic CRs. We have used the total inelastic cross sections, $\sigma_{inel}$, derived with the FLUKA Monte Carlo nuclear code \footnote{http://www.fluka.org/fluka.php}, which includes many channels for secondary particle production, in particular at low CR energies, and which has been benchmarked against accelerator data \citep{TorreLuque22}. We have calculated the interaction timescale of CR protons at both ends of the 5 to 500 GeV band.  
We obtain $\tau_{pp} n_H = a_{pp} = 36.5$ and 31.9~Myr~cm$^{-3}$ at 5 and 500 GeV, respectively. Heavier CR nuclei have shorter interaction times.

One can relate the spatial integral of equation \ref{eq:Lgam} to the total power, $\mathcal{P}_{CR}$, of CRs in the galaxy by introducing the average effective density of gas atoms encountered by the CRs as they diffuse through the gas:
\begin{equation}
    <n_{eff}> = \frac{\int_V n_H(\mathbf{r}) e_{CR}(\mathbf{r}) dV}{\int_V  e_{CR}(\mathbf{r}) dV}
\end{equation}
\noindent and by noting that the total energy of CRs in the galaxy can be expressed as:
\begin{equation}
    \int_V  e_{CR}(\mathbf{r}) dV = \mathcal{P}_{CR} \, \tau_{res} 
\end{equation}
\noindent so, in the calorimetric limit where $\tau_{res} \approx \tau_{pp}$, the \g-ray luminosity in equation \ref{eq:Lgam} becomes:
\begin{equation}
    L_{\gamma,cal} \approx  \frac{L_{\gamma H}}{e_{CR\odot}} <n_{eff}> \mathcal{P}_{CR} \tau_{pp} \approx \frac{L_{\gamma H}}{e_{CR\odot}} \mathcal{P}_{CR} \, a_{pp} \, .
\end{equation}
\noindent If CRs are primarily accelerated by SN shock waves and re-acceleration along their interstellar journey does not contribute much to their energy budget, the total CR power depends on the SN rate, $\Gamma_{\textrm{SN}}$, on the kinetic energy released per explosion, $E_{\textrm{SN}} = 10^{51}$~ergs, and on the $\eta_{\textrm{CR}}=10$\% fraction of this energy that is imparted to CRs. Uncertainties on this fraction are discussed in section \ref{sec:feedback}. One obtains $\mathcal{P}_{CR} = \Gamma_{\textrm{SN}} E_{\textrm{SN}} \eta_{CR}$ and, since the SN rate relates to the SFR in our simulations according to $\Gamma_{\textrm{SN}} = \eta_{\textrm{SN}} \textrm{SFR}/m_{\textrm{SN}}$, we find a calorimetric limit to the \g-ray luminosity which scales approximately as:
\begin{equation}
\label{eq:Lcal}
    L_{\gamma,cal} \approx \frac{\eta_{\textrm{SN}}}{m_{\textrm{SN}}}\, \mathrm{SFR} \,E_{\textrm{SN}} \, \eta_{CR}  \frac{L_{\gamma H}}{e_{CR\odot}} \, a_{pp} \, .
\end{equation}
Being proportional to the SFR through the SN rate, it appears as horizontal bands in Fig. \ref{Fig:gammaLum}.

The calorimetric limit in equation \ref{eq:Lcal} overlooks the different energy dependence of the hadronic and diffusive timescales. For instance, for a standard $3 \times 10^{28} (E_{k}/1\, {\rm GeV})^{0.3-0.6}$~\cmsqs diffusion coefficient, the hadronic timescale approaches the time is takes to diffuse to the 2-kpc scale height of the WIM for energies around 1 GeV, but it rapidly exceeds the diffusive timescale at lower as well as at higher energies: below 0.5 GeV because of the sharp drop in inelastic cross section, and beyond a few GeV because the diffusion coefficient increases whereas the inelastic cross section saturates. So CRs outside the one-to-few-GeV range can easily escape a galaxy unnoticed. This example illustrates why most of the galaxies in Fig. \ref{Fig:gammaLum} are poor calorimeters, except (maybe) for gas-rich, starburst galaxies. A more detailed exploration of the calorimetric efficiency of galaxies along the main sequence is discussed in \cite{Crocker21a}. The good calorimetric efficiency of starburst galaxies suggested in Fig. \ref{Fig:gammaLum} and the fact that they exhibit hard, SFR-independent spectral indices near 5~GeV in \g rays (${\sim}50$~GeV in CRs) in Fig. \ref{Fig:specgam} provides a consistent picture  if the diffusion coefficient of these particles is small enough to prevent significant leakage and if it has a weak dependence on energy so that the emitted radiation is closer to the CR injection spectrum than in the Milky Way. \cite{Krumholz20} have indeed estimated that CR self-streaming should prevail over most of the starburst medium and that sub-TeV particle should diffuse via the random walk of magnetic field lines with small ($10^{27-28}$\cmsqs), energy-independent diffusion coefficients.

\begin{figure}
   \centering

      \includegraphics[width=\linewidth]{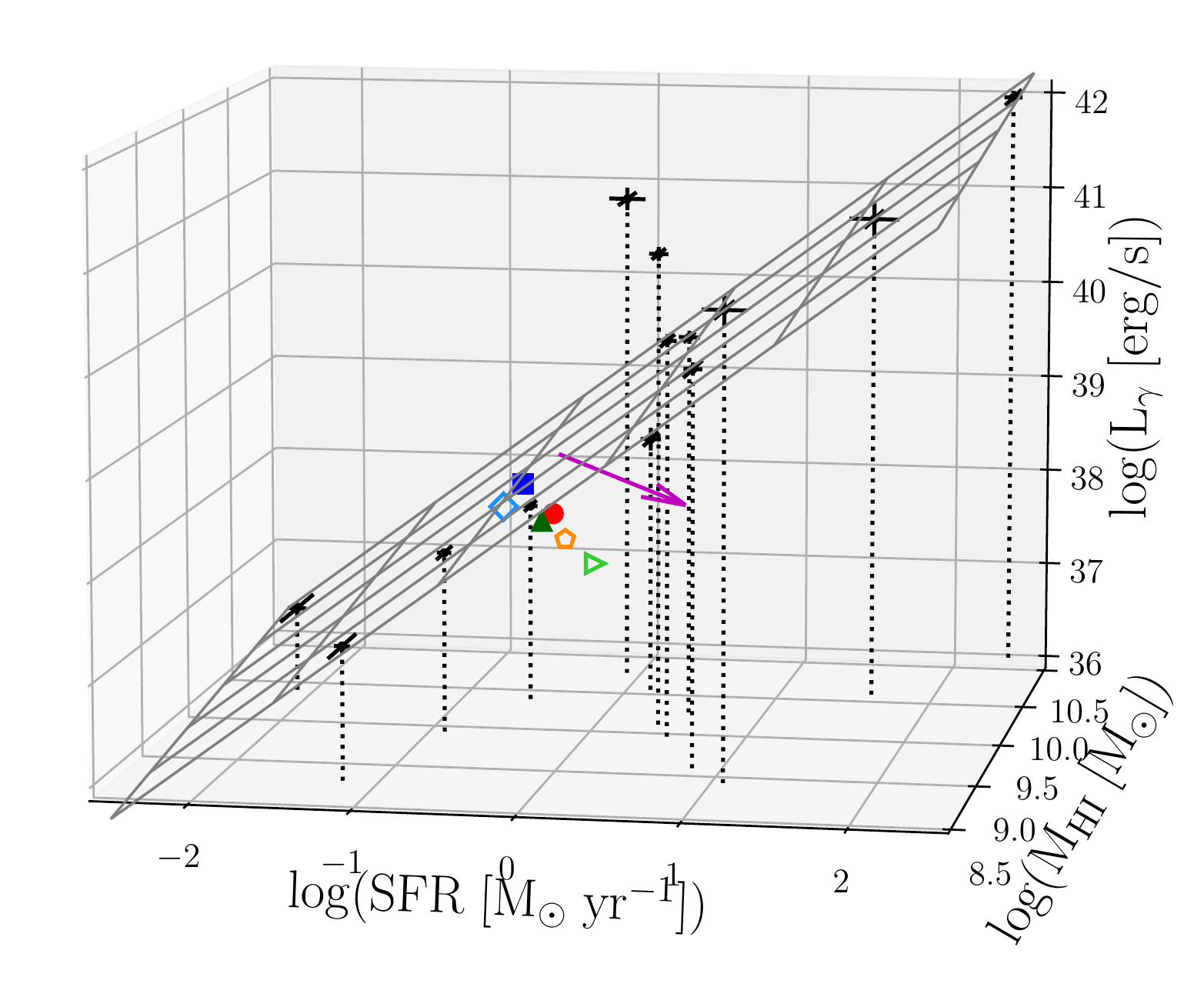}

      \caption{Evolution of the \g-ray luminosities, $L_{\gamma}$, in the 0.5-50 GeV band as a function of the SFR and atomic gas mass of the galaxy, showing how CR transport impacts the location of our simulated galaxies with respect to the best-fit plane followed by the Fermi LAT data points. The markers are the same as in Fig. \ref{Fig:gammaLum}. The arrow notes the perpendicular to the plane.}
         \label{Fig:gamplane}
   \end{figure}

 \begin{figure}
   \centering

      \includegraphics[width=\linewidth]{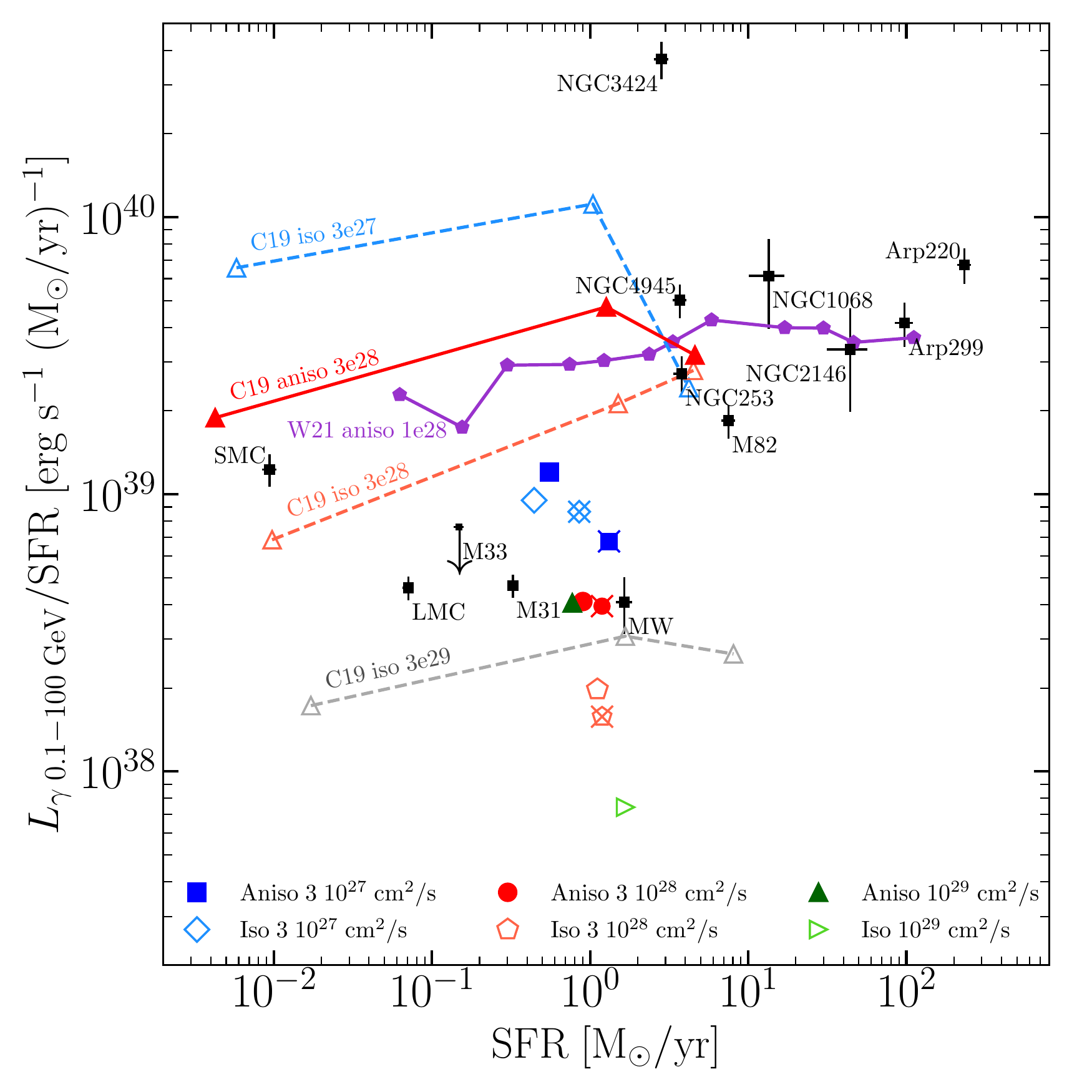}

      \caption{Correlation between the \g-ray luminosities in the 0.1-100 GeV band and the SFRs as obtained with Fermi LAT observations and  different simulations. The gas-rich dwarf galaxies simulated in this work and by \cite{Dashyan20} are shown as in Fig. \ref{Fig:gammaLum}. Results from \cite{Chan19} are given with triangles and colour coded according to $\kappa$ ($3\times 10^{27}$~\cmsqs in blue, $3\times 10^{28}$ in red or orange, and $3\times 10^{29}$ in grey). Results from \cite{Werhahn21} for $\kappa = 10^{28}$~\cmsqs are plotted as violet pentagons. Solid/dashed lines connect filled/open markers for anisotropic/isotropic  diffusion, respectively. All simulations use the same $\eta_{CR}=10$\% efficiency to inject SN energy into CRs. The \cite{Chan19} and present simulations model the multiphasic structure of the ISM whereas those of \cite{Dashyan20} and \cite{Werhahn21} use smoother monophasic gas.}
         \label{Fig:simcomp}
  \end{figure}

Figure \ref{Fig:gammaLum} shows that, even though the SFR in our simulated galaxies steadily rises with increasing $\kappa$ by a factor of nearly three, hence the injected CR power rises by the same amount, the global \g-ray luminosity drops by a factor of five, inducing a 16-fold drop in luminosity per SFR unit. This is not due to a change in target gas mass which we have checked to vary little between the different galaxies (by a few percent in the CNM+LNM or WNM phases, and by 20\% in the WIM). The marked evolution of the \g-ray luminosity with $\kappa$ reflects different residence times across the different ISM phases. 
As expected, Fig. \ref{Fig:gammaLum} shows that the global \g-ray luminosity is reduced for faster diffusion, and particularly so for isotropic diffusion, because the CRs leak out more rapidly into the low-density parts of the ISM and into the halo. The diffusion speed has less impact for anisotropic diffusion as the magnetic field configuration maintains the particles longer in the gas-rich environments. This is why the three anisotropic cases yield \g-ray luminosities within a factor of two despite a thirty-fold change in $\kappa$. They are all quite consistent with the \g-ray-SFR trend seen in the observations and we note that anisotropic diffusion with the canonical $\kappa$ value of $(3-10) \times 10^{28}$~\cmsqs is fully compatible with the expectations from M31 and the Milky Way.  The CR flux diverges more rapidly to the diffuse ISM for isotropic diffusion. The \g-ray luminosity for $\kappa = 3 \times 10^{28}$~\cmsqs deviates from the fitted trend at a $2\sigma$ level and we find that faster isotropic diffusion is not consistent with the trend despite the large dispersion in the observation sample. 

Interestingly, both panels of Fig. \ref{Fig:gammaLum} show that changing the mode of CR transport modifies the \g-ray luminosity per SFR or SFE unit almost perpendicularly to the galaxy-to-galaxy trend. This has been verified in the 3D (SFR, gas mass, \g-ray luminosity) space in Fig. \ref{Fig:gamplane} where, except for NGC 7059, the LAT data points closely delineate a plane. The simulated galaxies deviate further and further away from this plane as $\kappa$ increases, almost perpendicularly to the plane. The best-fit plane indicates that the \g-ray luminosity scales as $L_{\gamma} \propto {\rm SFR}^{1.19 \pm 0.26} M_{\rm HI}^{0.31 \pm 0.56}$, therefore primarily linearly with the SFR and hardly with the gas mass available for CR interactions. A larger sample is needed to verify if the gas mass has a small influence or none. In any case, the Fermi LAT data can be used to test simple CR transport assumptions, as earlier proposed by \cite{Pfrommer17} and \cite{Chan19}, but with care since the contribution to the total galaxy flux of numerous internal point sources such as pulsars and SN remnants cannot be resolved, except with difficulty in the LMC \citep{Ackermann16}. 

\subsection{Comparison with other simulations}
\label{sec:simcomp}

The results of the FIRE-2 simulations by \cite{Chan19} and \cite{Hopkins21gam} led them to conclude that large transport speeds with $\kappa$ of order $10^{29-31}$~\cmsqs are necessary to explain the observations. Our results do not support this conclusion, to the contrary, as we find good consistency with the observations for anisotropic $10^{27.5-29}$~\cmsqs diffusion and for isotropic diffusion with $\kappa \lesssim 3 \times 10^{28}$~\cmsqs. Faster isotropic diffusion is beyond 2.5 $\sigma$ from the observation trend. We also find systematically fainter \g-ray luminosities than the results of the AREPO simulations performed by \cite{Werhahn21}. Figure \ref{Fig:simcomp} compares the \g-ray yields of the different simulations in the broader 0.1-100 GeV \g-ray band, for isolated galaxies and for the same CR injection efficiency, $\eta_{CR}{=}10$\%. It shows that the FIRE-2 luminosities are typically ten times brighter than ours for the same diffusion coefficient, whether CRs propagate isotropically or along magnetic field lines. This ratio is reduced to a factor around 5 if we compare with the AREPO simulations, even though their results nicely match the Fermi LAT data at large SFR. 

We have noticed that the \g-ray emissivity used above 1 GeV in FIRE-2 per gas atom, when scaled to the CR energy density near the Sun, is 2.4 times larger than what is actually measured with Fermi LAT in the solar neighbourhood. Correcting this over-emissivity is, however, not sufficient to explain the order of magnitude difference with our simulations. They also use a power-law CR spectrum with a spectral index of 2.2 that is close to what is possibly injected by SN shock waves, but at variance with the much softer spectrum that is observed in the Milky Way and the LMC after propagation in the ISM (see Fig. \ref{Fig:specgam}). \cite{Werhahn21} also inject an $E^{-2.2}$ spectrum, but they follow the energy-dependent propagation of the particles and their spectrum softens after propagation (see their Fig. 7). Nevertheless, we have checked that these spectral differences would change the integrated \g-ray luminosity by only ${\pm}50$\% when normalised to the CR energy density $e_{CR\odot}$. This is not enough to explain the differences seen between the simulations. A detailed modelling of energy-dependent CR transport is central to modelling galaxy spectra, but much less so to study the \g-ray-FIR trend. 

All the simulation results shown in Fig. \ref{Fig:simcomp} use simple uniform diffusion in addition to CR advection by the gas, so different solutions to ISM-dependent CR transport properties cannot be advocated to explain the disparities. The codes use different approximations to solve the CR transport equation, but \cite{Chan19} have checked in their Appendix B3 that the impact on the SFR and \g-ray luminosities is small. \cite{Thomas2022} also conclude that solving the two-moment expansion might be more important for CR transport in the winds, but not necessarily in the ISM, since the steady-state approximation holds within the galactic disc.

The disparity may stem from different gas dynamics. The spatial distribution and the clumpiness of the target gas for hadronic interactions are largely dictated by the thermal state of the gas. Both the \cite{Chan19} and the present simulations have modelled the highly-contrasted structure of multiphasic gas whereas those by \cite{Werhahn21} and \cite{Dashyan20} describe smoother, monophasic gas. Since the present work uses the very same code, same maximum resolution, and same initial galaxy set-up as in \cite{Dashyan20}, we can compare them to gauge the impact of the multiphasic or monophasic description of the ISM. Figures \ref{Fig:gammaLum} and \ref{Fig:simcomp} show that the \g-ray luminosity per SFR or SFE unit decreases slightly for the smoother gas simulations, likely because the particles encounter lower gas densities in the early part of their interstellar journey. We find a modest decrease by less than 30\%, except for the very slow ($3\times 10^{27}$~\cmsqs) anisotropic case for which the difference reaches a factor of two. In this case, the CRs flow along magnetised, CNM-rich filaments (i.e. spurs and arms in Fig. \ref{Fig:EnergyMapsFaceon}) and it is necessary to adequately describe the density contrast along their journey to correctly model the \g-ray production. For faster and/or less guided transport, the impact of the gas description on the \g-ray yield is small because the typical scales of CR gradients are much larger than those of CNM structures (see Fig. \ref{Fig:EnergyMapsFaceon}). We will further explore the impact on the \g-ray yield at cloud scales in the companion paper. Nevertheless, describing a more contrasted ISM in the AREPO simulations may increase rather than decrease the total \g-ray luminosity of a galaxy if it follows the same trend as found here. The comparison anyhow indicates that the monophasic or multiphasic description of the gas should affect the global \g-ray luminosity of simulated galaxies by less than a factor of two over this range of diffusion coefficients. 

The magnetic-field distribution also impacts the gas structure by changing the pressure support against gravitational collapse and by directing mass transfers \citep{Hennebelle13}. Planck data on dust polarisation shows that dense gas tends to form filaments perpendicular to the magnetic field, suggesting that gas accumulation and gravitational collapse occur along the magnetic field lines \citep{Planck16}. This is only possible if the turbulence is \alfvenic or sub-\alfvenic, implying that the magnetic field plays a significant role in the dynamics and structuring of the gas from the WNM to the CNM phase and inside the latter. We find magnetic-field strengths across the simulated galactic discs that compare well with ISM measurements and with the lower-resolution results of \cite{Werhahn21}. Figure \ref{Fig:Pratios} shows that, on scales of hundreds of parsecs, the magnetic pressure often dominates over the other terms. On the contrary, Figure 5 of \cite{Hopkins2020b} shows that the magnetic pressures in the simulated galaxies are orders of magnitude below the thermal and CR pressures. We cannot compare their $10^{11}$~M$_\odot$ dwarf (m11b) with ours as it forms stars at a much lower rate of 0.01 M$_\odot$/yr. With less pressure support, one would expect a larger mass fraction in their dense gas phases than in our simulations, thereby enhancing the \g-ray production in the vicinity of CR sources, but this needs be verified by comparing similar dwarfs. 
We could not shed further light on the over-brightness of the FIRE-2 and AREPO simulations compared to ours and compared to the Fermi LAT observations around Milky-Way-like rates of star formation.

\section{Conclusions }
\label{sec:Conclusions}

We have performed high-resolution (9 pc) MHD simulations of an isolated disc galaxy with a total mass of  10$^{11}$~M$_{\odot}$ using the RAMSES code. These simulations include CR injection and two simple modes of CR transport in addition to particle advection by the gas: one mode where CRs diffuse isotropically from their sources and one where they diffuse preferably along the magnetic field lines. We have tested different diffusion coefficients ranging between $3\times 10^{27}$ and $10^{29}$~\cmsqs in both modes. The resulting spatial distributions of CRs significantly change across the galaxies. Isotropic diffusion generates smoother distributions with localised over-densities around the star forming sites whereas anisotropic transport concentrates CRs along the spiral arms and spurs. Both CR distributions broaden radially and in height above the galactic plane as the particles are allowed to escape more rapidly from their injection sites. 
Our simulations include gas cooling and heating prescriptions that allow to model the multiphasic structure of the ISM and to study global effects that CRs might exert on it. We have focused this first article on the large-scale properties of the different components and on the global SFR and \g-ray luminosity of the galaxies and we summarise below the main results.

\begin{itemize}

\item \textbf{Simple CR diffusion weakly modifies the large-scale spatial distribution of the gas in the galaxy.} The radial distribution of the gas is insensitive to the modes of CR propagation that we have tested. The thickness of the gaseous disc does not significantly change when we vary the diffusion speed along the magnetic field because the CRs remain largely confined along the toroidal field in the disc. Conversely, the gas hydrostatic equilibrium responds to changes in vertical CR pressure gradients in the case of isotropic diffusion, doubling its thickness when the diffusion speed is reduced by thirty. This swelling occurs primarily in the diffuse WIM, the thickness of the denser WNM varying by only 30\%-40\%. The radial and vertical distributions of the dense gas phases (CNM and LNM) are particularly stable against CR changes. 
   
\item \textbf{CR transport modifies the large-scale magnetic field distribution in the galaxy.} Given our choice of SN feedback parameters, the turbulent and CR pressures generally dominate over the thermal and magnetic ones on scales of 100-200~pc across the whole galaxy. Their radial profiles are closely related except when the particles are allowed to rapidly and smoothly diffuse outward. On those scales, the mean magnetic-field strengths approach equipartition values in the inner regions and remain a factor around five below equipartition elsewhere. We observe a significant CR feedback on the amplification of the magnetic field in the inner half of the galaxies. The mean strengths are increased by a factor of 3.5 for anisotropic diffusion and for slow isotropic propagation, hence when the CR energy density builds up beyond ${\sim}2$~eV cm$^{-3}$. Given the highly-contrasted filamentary structure of the magnetic-field distribution, we will further investigate this amplification in the companion paper to study how it couples to turbulence, CR gradients, and to vertical outflows in magnetised fountains.

\item \textbf{The global mass ranking in the different gas phases varies little with CR transport and compares reasonably well with estimates in the Milky Way}, so does the ratio of the stable and unstable phases in the total neutral gas. The comparison with the Milky Way should be considered with care because of the size and mass difference with the simulated dwarfs, but the gas circulation between the phases should not dramatically differ in these gas-rich dwarfs from what we observe in the Milky Way since they produce stars at an equivalent rate or only two to three times less. Inferences of the mass fractions in the Milky Way phases also lack precision and all the simulated results lie well within the observation uncertainties. Changes in CR transport induce less than 5\% variations in the mass of the main WNM reservoir and a 20\% increase in CNM mass with increasing diffusion speed in the isotropic case only. Smaller scale studies, at the scale of spiral arms, cloud complexes and individual clouds, will be presented in the companion paper to describe the effect that CR dynamics have on the gas circulation between the different phases of the ISM, on the structure of the gas, and on its thermal distribution.
   
\item \textbf{CR transport has a dual impact on star formation}. CR dynamics directly help regulate star formation by providing additional pressure support to dense clouds against gravitational collapse while their pressure gradients at intermediate (0.1-1 kpc) scales limit the gas circulation from the large WNM reservoirs to the CNM cloud precursors to star-forming molecular clouds.  
But CR feedback also acts indirectly on star formation via magnetic amplification. We have shown that star formation is suppressed by an average factor up to 2.5 for anisotropic or for slow isotropic diffusion compared to isotropic propagation with $\kappa \geq 3 \times 10^{28}$~\cmsqs. The suppression primarily relates to a twofold to threefold increase in mean magnetic pressure as well as mean CR pressure in the CNM phase in the inner half of the galaxies. The same increase in CR pressure in the outer regions is not associated with a similar increase in magnetic pressure and it has a smaller impact of order 50\% on the local SFR. Smaller scale studies will shed more light on the efficiency of CR feedback on the magnetic field strength in different environments, hence on the importance of CR-induced magnetic feedback on SFR regulation in more active types of galaxies. 
   
\item \textbf{Fast CR diffusion is disfavoured by \g-ray and SFR observations}. We have calculated the hadronic \g-ray luminosities of the simulated galaxies in the 0.5 to 50 GeV to reduce contamination by CR electron radiation and we have updated the data from the sample of galaxies detected by Fermi LAT so far. The LAT sample is still too sparse to test whether the total gas mass of a galaxy influences the \g-ray luminosity in addition to the SFR or not. It is unfortunate as a dual trend would tighten the constraints one can derive on CR transport from the \g-ray-SFR relation.

For the simulations, we have used the \g-ray emissivity measured in clouds around the Sun, which corresponds to the total CR energy density surrounding the heliosphere for CR nuclei above 1 GeV/nucleon. We have assumed a uniform CR spectrum across the galaxies. This is motivated by the uniformity of the hadronic emission spectra recorded in a large variety of environments in the Milky Way. Changes by ${\pm}0.1$ in spectral index, commensurate with the amplitude of the possible spectral gradient found radially across the Milky Way, would modify the broad-band \g-ray luminosities by ${\pm}15$\% only. 

We find that the \g-ray and SFR yields of the simulated galaxies are fully consistent with the best-fit trend seen in the observations in the case of anisotropic $10^{27.5-29}$~\cmsqs diffusion and for isotropic diffusion with $\kappa \leq 3 \times 10^{28}$~\cmsqs. Good consistency is found with M31 and the Milky Way, close in SFR to the simulated dwarfs. The total \g-ray luminosity departs below the observation fit as the diffusion coefficient increases because CRs rapidly escape out of the denser phases where they are more prone to undergo hadronic interactions with the ambient gas. We find isotropic diffusion with $\kappa \geq 1 \times 10^{29}$~\cmsqs to be inconsistent with the observations.  

This conclusion is in tension with the results of other simulations using the same CR transport modes for isolated galaxies \citep{Chan19} and cosmological simulations \citep{Hopkins21gam}, where they report the need for $10^{29-31}$~\cmsqs diffusion coefficients to reproduce the observations. Comparing simulated galaxies that have evolved in a cosmological environment is not trivial since CRs modify the gas circulation in the circum-galactic medium, thus impact the accretion history and growth of the disc \citep{Buck2020}. Restricting the comparison to isolated galaxies and to the same choices of CR transport and of CR power injection, we find that the galaxies simulated by \cite{Werhahn21} with AREPO are about five times brighter per SFR unit than ours for the same CR propagation mode, and this ratio increases to ten for the galaxies simulated by \cite{Chan19} with FIRE-2. As discussed at the end of section \ref{sec:gamma}, those disparities cannot be readily attributed to different approximations in the CR transport solver, to different assumptions in CR spectrum, or to the monophasic or multiphasic modelling of the gas. We have noted the use in FIRE-2 of a \g-ray emissivity per gas atom that exceeds the Milky Way measurements by a factor of 2.5 for the same CR energy density, but such a correction is not large enough to reconcile their predictions with ours. 
   
\end{itemize}

These disparities illustrate that the observed trend between the Fermi LAT data and SFR estimates derived from FIR emission or other tracer combinations will usefully constrain CR propagation properties, but in a way that is obviously not fully understood yet. Finding significant tensions for the simplest case of CR advection by the gas and uniform diffusion illustrates the complexity of the problem at hand. It suggests that the difference may stem from the overall dynamics of the gas, hence on its thermal and magnetic state. The coupling between CR dynamics and the magnetic field apparently plays a prime role on the final gas state and SFR. The data from our simulations corroborate the AREPO findings \citep{Pakmor16} and we will further explore this coupling at sub-galactic scales. The origin of the differences must be clarified to allow reliable tests of more complex CR propagation schemes in the future. This is why, rather than extending our simulations along the main sequence of galaxies, from dwarfs to starbursts, we will use the present dwarfs with uniform diffusion coefficients as a benchmark to compare the impact of ISM-dependent diffusion properties in the next runs.

\begin{acknowledgements}
We acknowledge the financial support by the LabEx UnivEarthS (ANR-10-LABX-0023 and ANR-18-IDEX-0001) for this work. This work was granted access to the HPC resources of IDRIS under the allocation 2022-A0090412046 made by GENCI.
\end{acknowledgements}

\bibliographystyle{aa}
\bibliography{biblio}

\begin{appendix} 

\section{Galaxy properties}
\label{sec:galprop}

\begin{table*}[h]
\caption{ Distance, atomic gas mass, SFR, $8{-}1000$~$\mu$m luminosity, and 0.5-50 GeV luminosity of the galaxies shown in Figure \ref{Fig:gammaLum}.}
\label{tab:gammaSFR}
\centering
\begin{tabular}{cccccc}
\hline
Galaxy  & \begin{tabular}[c]{@{}c@{}}M$_{\textrm{gas}}$ \\ {[}$10^{9}$ M$_{\odot}${]}\end{tabular} & \begin{tabular}[c]{@{}c@{}}distance\\ {[}Mpc{]}\end{tabular} & \begin{tabular}[c]{@{}c@{}}SFR\\ {[}M$_{\odot}$ yr$^{-1}${]}\end{tabular} & \begin{tabular}[c]{@{}c@{}}$L_{\textrm{IR}}$ \\ {[}$10^{10}\;L_{\odot}${]}\end{tabular} & \begin{tabular}[c]{@{}c@{}}$L_{\gamma}$ \\ {[}$10^{38}\;$erg s$^{-1}${]}\end{tabular} \\ \hline
MW & $8.0\pm 1.6$ & - & $1.65 \pm 0.19$ & - & $4.20 \pm 0.84$ \\
M31     & $5.4 \pm 0.8$ & $0.78 \pm 0.02$ & $0.33 \pm 0.02$ & $0.24 \pm 0.02$ & $1.02 \pm 0.06$ \\
M33     & $1.9 \pm 0.3$ & $0.81 \pm 0.02$ & $0.15 \pm 0.01$ & $0.110 \pm 0.008$ & $<0.66 $         \\
M82     & $0.9 \pm 0.2$ & $3.53 \pm 0.25$ & $7.5 \pm 0.7$ & $5.60 \pm 0.6$ & $88.0 \pm 8.8$ \\
LMC     & $0.5 \pm 0.1$ & $0.050 \pm 0.002$ & $0.071 \pm 0.006$ & $0.053 \pm 0.004$ & $0.21 \pm 0.01$ \\
SMC     & $5.6 \pm 1.9$ & $0.060 \pm 0.004$ & $0.009 \pm 0.001$ & $0.007 \pm 0.001$ & $0.072 \pm 0.006$ \\
NGC 253 & $4.2 \pm 0.6$ & $3.56 \pm 0.25$ & $3.8 \pm 0.4$ & $2.8 \pm 0.3$ & $71.4 \pm 7.1$ \\
NGC 1068 & $0.6 \pm 0.2$ & $10.1 \pm 1.8$ & $13.5 \pm 3.4$ & $10.1 \pm 2.5$ & $419 \pm 106$  \\
NGC 2146 & $8.6 \pm 2.6$ & $27.7 \pm 5.6$ & $44.2 \pm 12.6$ & $32.9 \pm 9.4$ & $862 \pm 247$  \\
NGC 3424 & $2.9 \pm 0.5$ & $27.2 \pm 2.0$ & $2.8 \pm 0.3$ & $2.1 \pm 0.2$ & $606 \pm 66$   \\
NGC 4945 & $2.1 \pm 0.4$   & $3.72 \pm 0.26$ & $3.7 \pm 0.4$ & $2.8 \pm 0.3$ & $99.4 \pm 9.9$ \\
NGC 7059 & $13.1 \pm 3.1$ & $31.9 \pm 4.5$ & $0.94 \pm 0.20$ & $0.70 \pm 0.15$ & $991 \pm 198$  \\
Arp 220 & $31.6 \pm 9.6$ & $84.3 \pm 6.0$ & $233 \pm 23$ & $173 \pm 17$ & $9089 \pm 946$ \\
Arp 299 & - & $48.6 \pm 4.0$ & $97.5 \pm 12.7$ & $72.6 \pm 9.4$ & $2156 \pm 258$ \\ 
\hline
\multicolumn{6}{c}{simulations from Dashyan \& Dubois 2020} \\ 
\hline
Aniso 327 & 1.24 & * & 1.32 & * & 5.91 \\
Iso 327 & 1.11 & * & 0.85 & * & 4.90 \\
Aniso 328 & 0.92 & * & 1.19 & * & 3.12 \\
Iso 328 & 0.95 & * & 1.19 & * & 1.25 \\ 
\hline
\multicolumn{6}{c}{simulations from this work} \\ 
\hline
Aniso 327 & 1.34 & * & $0.55 \pm 0.05$  & * & $4.42 \pm 0.51$   \\
Iso 327 & 1.19 & * & $0.44 \pm 0.06$  & * & $2.80 \pm 0.47$   \\
Aniso 328 & 1.21 & * & $0.90 \pm 0.07$  & * & $2.45 \pm 0.50$   \\
Iso 328 & 1.10 & * & $1.11 \pm 0.13$  & * & $1.46 \pm 0.32$   \\
Aniso 129 & 1.18 & * & $0.77 \pm 0.07$  & * & $2.08 \pm 0.45$   \\
Iso 129 & 1.15 & * & $1.68 \pm 0.16$  & * & $0.83 \pm 0.11$   \\ 
\hline
\end{tabular}
\end{table*}

Table \ref{tab:gammaSFR} lists the properties of the galaxies used in Figure \ref{Fig:gammaLum} for the real objects as well as the simulated ones. For the former, we paid attention to preserving the uniformity of the sample as much as possible. All distances, $D$, correspond to Tully-Fisher estimates obtained in the CosmicFlows4 survey \citep{Kourkchi20}. All SFRs have been derived from the broadband 8-1000 $\mu$m FIR luminosities, following the prescription given by \cite{Sanders96} to combine the fluxes recorded in the four IRAS bands: 
\begin{equation}
F_{FIR} = 1.8 \times 10^{-14} \times (13.48 f_{12} + 5.16 f_{25} + 2.58 f_{60} + f_{100})~{\rm W~m}^{-2} \end{equation}
\noindent where $f_{12}$, $f_{25}$, $f_{60}$, and $f_{100}$ note the IRAS flux densities (in Jy) at 12, 25, 60, and 100 $\mu$m, which come from the \cite{Sanders03} and \cite{Moshir90} source catalogues.
The FIR luminosities have been converted to SFRs according to \cite{Kennicutt1998}:
\begin{equation}
    \textrm{SFR} = 1.7 \times 10^{-10} \, \epsilon \, \left(\frac{\textrm{L}_{8-1000\mu\textrm{m}}}{\textrm{L}_{\odot}}\right) \textrm{M}_{\odot}\; \textrm{yr}^{-1}
\end{equation}
\noindent where the $\epsilon=0.79$ factor corresponds to the \cite{Chabrier2003} IMF which has been used for star formation in our simulations. 
The derived SFRs compare reasonably for the purpose of Fig. \ref{Fig:gammaLum} with estimates exploiting different sets of tracers, for instance combining H$\alpha$ and FIR information \citep[e.g. 0.2 and  0.036-0.1 M$_{\odot}$ yr$^{-1}$ for the LMC and SMC, respectively,][]{Harris09}, or UV and 25 $\mu$m fluxes \citep[e.g. $0.20 \pm 0.03$ and $0.027 \pm 0.003$ M$_{\odot}$ yr$^{-1}$ for the LMC and SMC, resp.,][]{Kornecki2020}, or the full galaxy SED \citep[$0.28^{+0.02}_{-0.01}$ M$_{\odot}$ yr$^{-1}$ for M33,][]{Thirlwall20}. The discrepancies, however, highlight that systematic uncertainties largely dominate over the measurement ones and that the quoted SFR values could be off by a factor of 2. 

Atomic gas masses have been estimated from \hi line flux integrals, $f_{21}$, retrieved from the Hyperleda database (http://leda.univ-lyon1.fr) and corrected for self absorption. They have been converted to gas mass according to \cite{Butcher16}: 
\begin{equation}
M_{\rm HI} = 2.36 \times 10^5 M_{\odot} (D/Mpc)^2 \, f_{21} .
\end{equation}
\noindent Atomic gas generally dominates the gaseous mass content of a galaxy. To preserve the homogeneity of the sample, we did not add molecular masses because the conversion ratios to transform molecular line fluxes into \hd mass have large uncertainties and large scatter from one galaxy to another, especially in extreme starburst environments or low metallicity dwarfs. Masses of dark gas equivalent to the CO-bright molecular mass can also be hidden from \hi and molecular line surveys at the \hi-to-\hd interface. The SFE presented in Fig. \ref{Fig:gammaLum} should therefore be considered as upper limits, of order 30\%-50\% above the actual values. 

The sample of \g-ray galaxies includes the latest associations proposed between Fermi LAT sources and nearby or starburst galaxy counterparts in the 4FGL-DR3 catalogue \citep{Abdollahi224FGLDR3}. The luminosities have been derived in the 0.5-50 GeV band from the spectral characterisation performed for each source in the catalogue, taking advantage of twelve years of continuous observation of the whole sky. As mentioned in the text, the detection of M33 has not been confirmed in 4FGL-DR3, despite the increased observation duration, so we list the $1\sigma$ upper limit obtained by \cite{Ackermann2017}. The LMC and SMC luminosities correspond to the detection of extended emission toward the galaxy. The flux does not include several point sources that could be resolved in the LMC. 


\section{Rotation curves}\label{App:rotationcurves}
\begin{figure*}[h]
   \centering
      \includegraphics[width=0.95\textwidth]{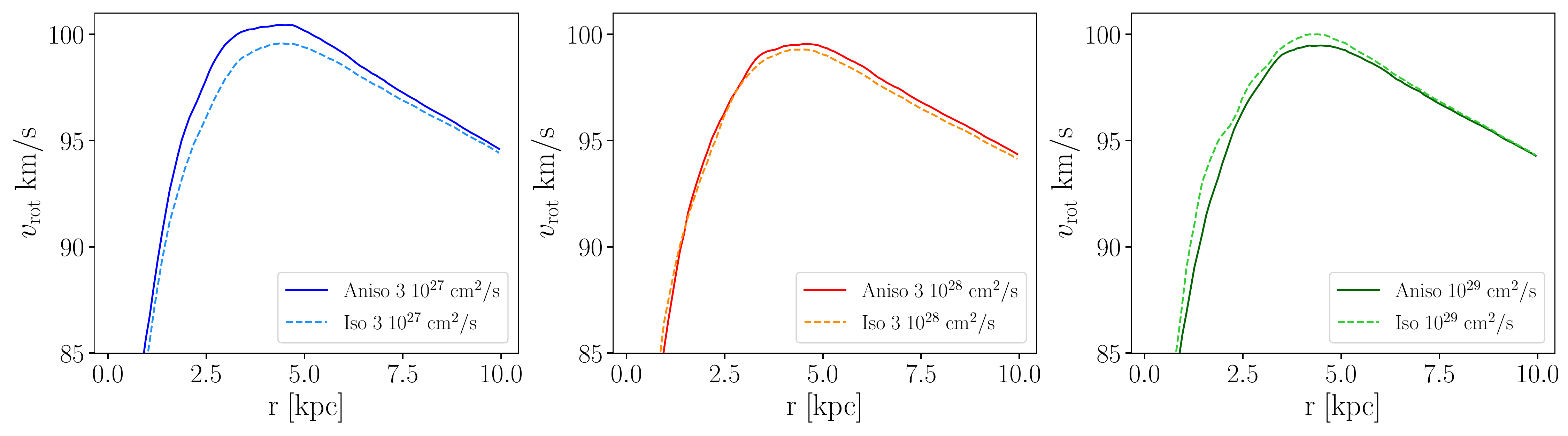}
      \vspace{-3\baselineskip}
      \vspace{2\baselineskip}
      \caption{Spherical rotation curves for the different runs, computed using all the massive components of the galaxy, dark matter, gas and stars. The curves are separated by diffusion parameters to avoid overcrowding.}
         \label{Fig:RC}
   \end{figure*}
In figure \ref{Fig:RC} we show the rotation curves (RC) of all the simulated galaxies in three panels, with the diffusion coefficient increasing from left to right. The RCs are computed using all the massive components of the simulation, i.e. stars, gas and DM as: 
\begin{equation}
    V_{\rm circ}(r) = \sqrt{\frac{G M_{\rm tot}(<r)}{r}},
\end{equation}
\noindent where $G$ is the gravitational constant, $M_{\rm tot}(<r)$ is the total mass inside the radius $r$. All curves are fairly similar, so we have chosen to separate them in three panels. The rotation curves peak at ~100 km/s at a radius close to 3 kpc.

\section{Alfv\'en Mach number} \label{App:Machnumbers}
\begin{figure*}
   \centering
    \begin{subfigure}{0.9\linewidth}
      \includegraphics[width=0.95\textwidth]{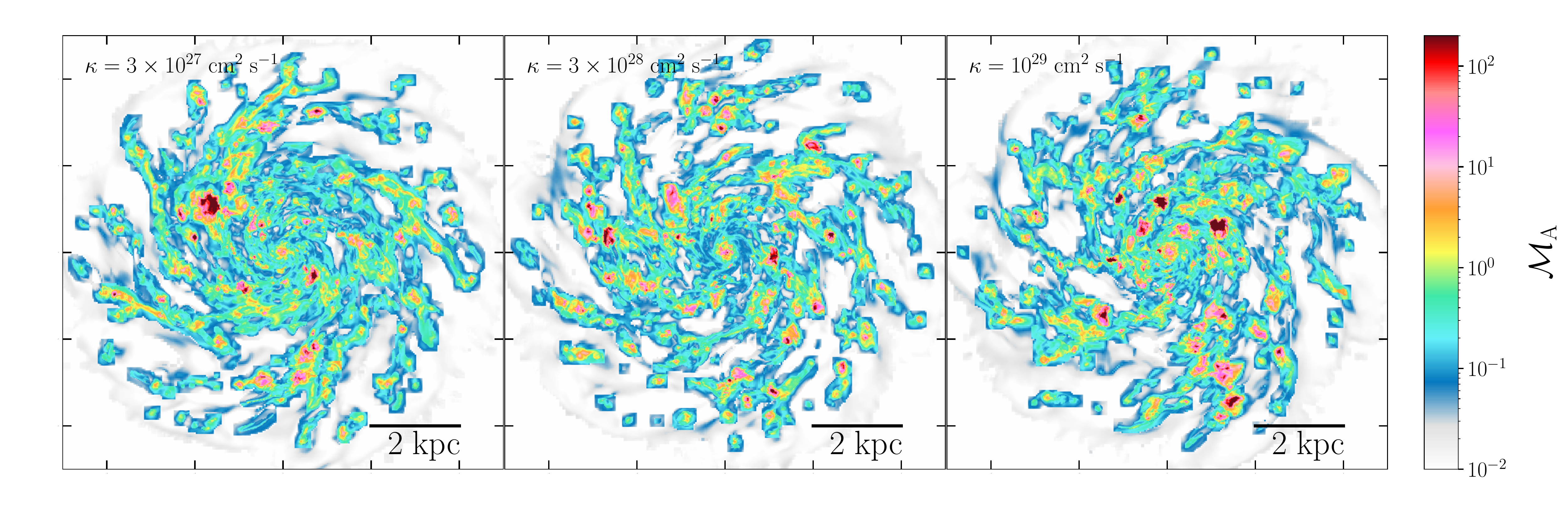}
      \caption{Anisotropic diffusion}
    \end{subfigure}
    \begin{subfigure}{0.9\linewidth}
      \includegraphics[width=0.95\textwidth]{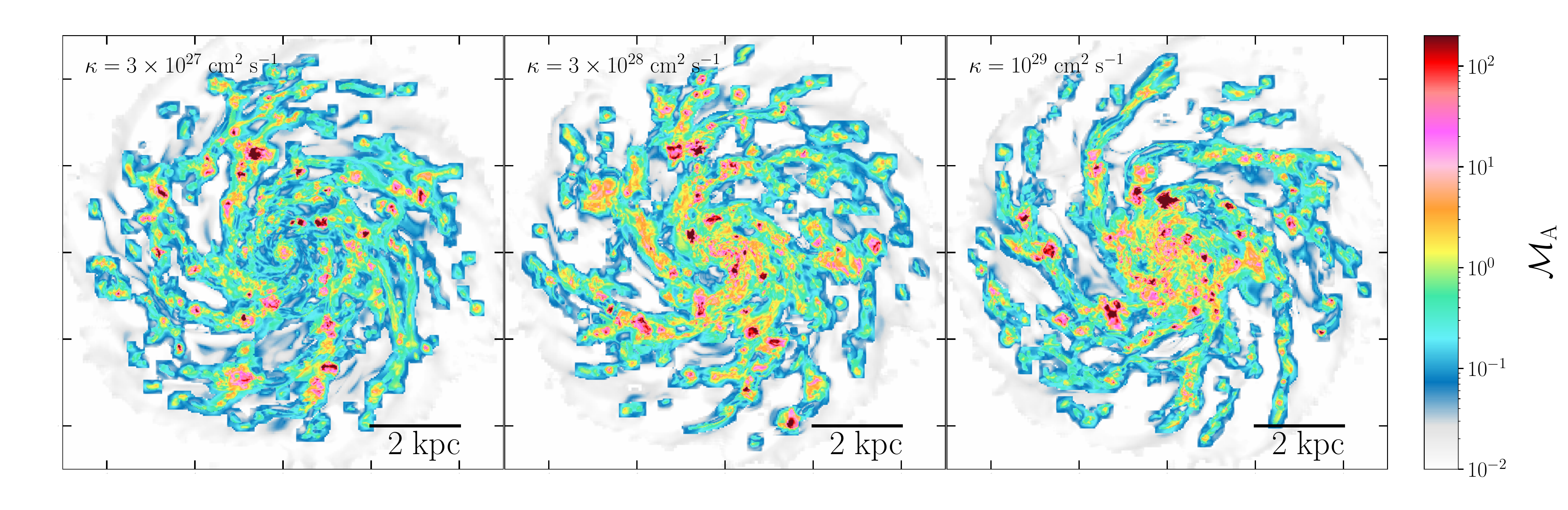}
      \caption{Isotropic diffusion}
    \end{subfigure}

      \vspace{-3\baselineskip}
      \vspace{2\baselineskip}
      \caption{Edge-on maps of the Alfv\'en Mach number for runs with anisotropic (top) and isotropic (bottom) CR diffusion, calculated for all gas within $\pm 50$~pc from the galactic plane.   }
         \label{Fig:MachA}
   \end{figure*}
In Figure \ref{Fig:MachA} we show the \alfven Mach number 
$\mathcal{M}_{\rm A}=v_{\rm turb}/v_{\rm A}$, as the ratio of the turbulent velocity and the ideal Alfv\'en velocity, $v_{\rm A} = |B|/\sqrt{4 \pi \rho}$. For the maps shown in Figure \ref{Fig:MachA} we use the gas that is within $\pm 50$~pc from the galactic plane. The \alfven Mach number distributions are very wide, showing maxima around 200 and minima near 0.1. The mean of the distributions vary with the CR transport mechanisms, the mean in the anisotropic transport case being generally lower than in the isotropic case. With growing diffusion coefficients, we obtain average \alfven Mach numbers of 2.1, 2.5, 2.8 for anisotropic runs and 2.8, 4.1 and 3.4 for isotropic runs.  

The increase in \alfvenic Mach numbers in the inner regions seen for increasing isotropic speed, but not for anisotropic transport, relates to the changes in magnetic-field strength and SFR discussed in section \ref{sec:CRandSFR}.
 
\section{Scale heights over time}
\label{sec:time}
\begin{figure}
   \centering
     \begin{subfigure}{0.8\linewidth}
      \includegraphics[width=\textwidth]{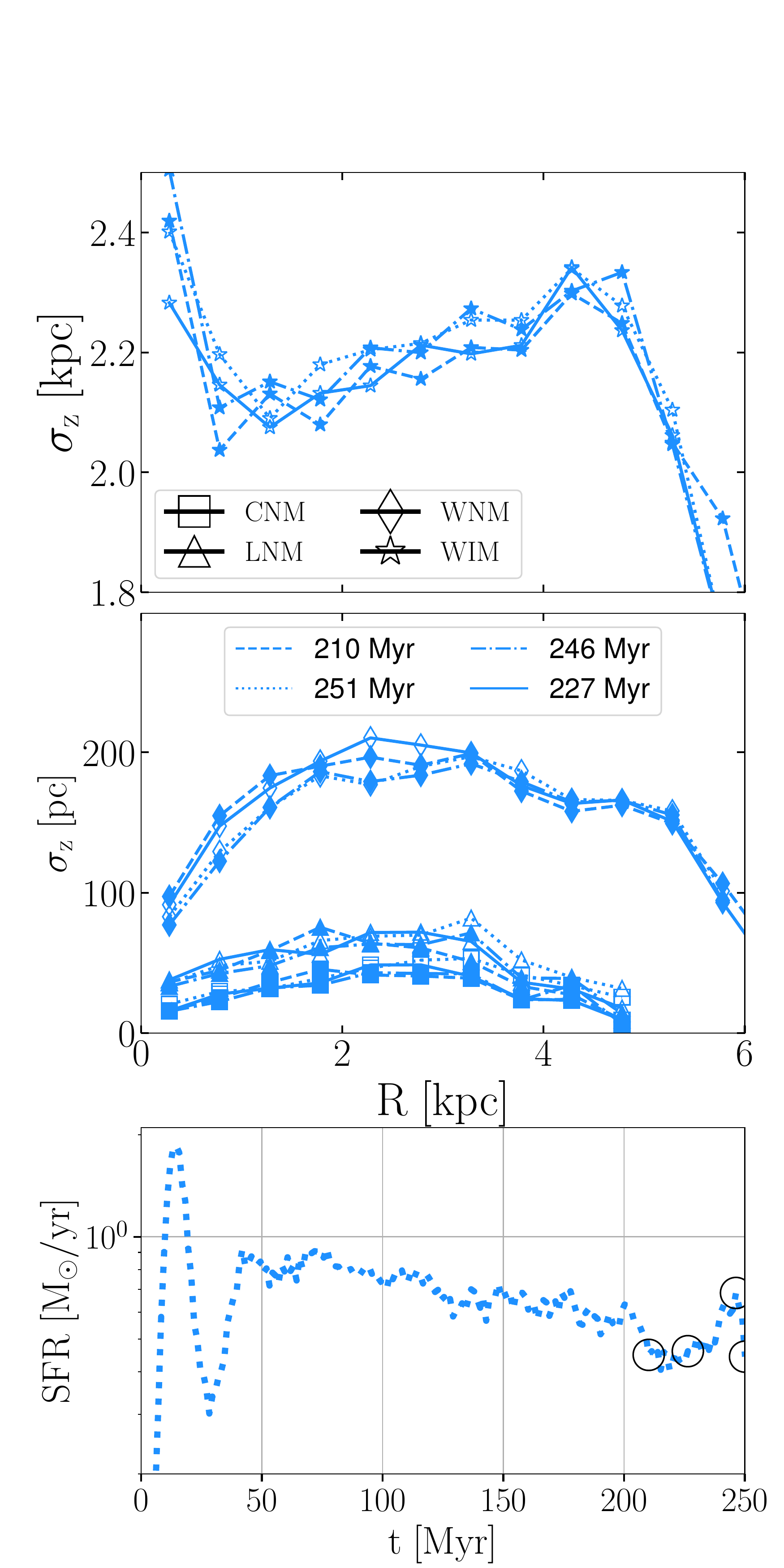}
   \end{subfigure}
    \begin{subfigure}{0.8\linewidth}
      \includegraphics[width=\textwidth]{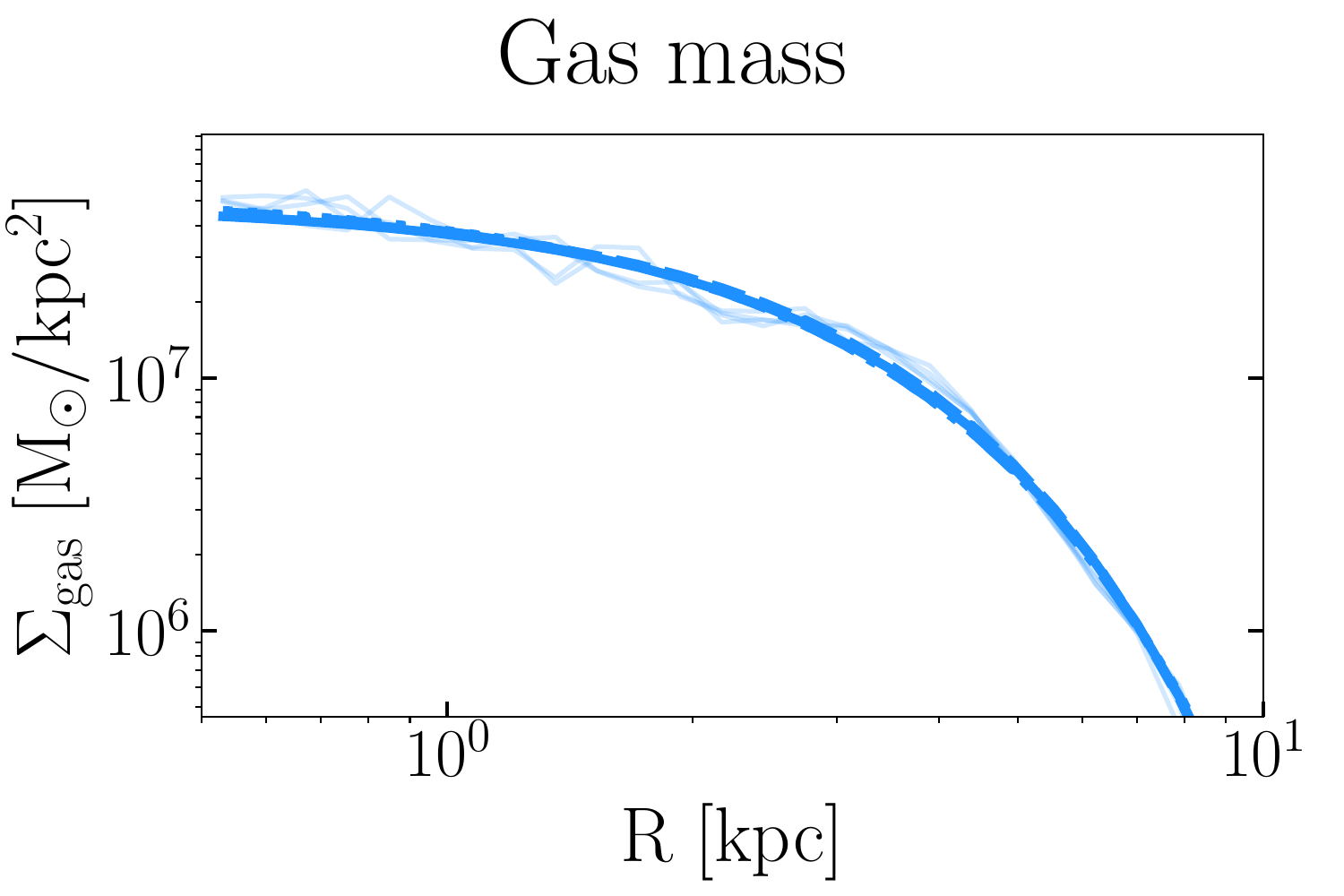}
   \end{subfigure}

 \caption{Time evolution of large-scale gas characteristics for the run with isotropic diffusion and $\kappa=3\times 10^{27}$~\cmsqs. The vertical scale height of the gas in the ISM phases is shown in the upper two panels as in Figure \ref{Fig:phasescaleheight}. The third panel shows the (circled) moments in history chosen for their different SFR levels to display the gas characteristics. The lower panel shows the radial profiles (faint curves) in gas mass surface density for the same moments in history. The individual S\'ersic curves fitted to these profiles (dark curves) pile up due to the time stability.}
         \label{Fig:sigmaztime}
   \end{figure}

In order to test the time evolution of the vertical dispersion, $\sigma_z$, of the gas in the different phases and the global radial profile of gas over time, we have calculated them at different moments of the galaxy evolution. More specifically, we chosen moments in history exhibiting different SFRs, at a peak or recession in star formation, or during an upturn and a downturn in activity. One example is shown in Figure \ref{Fig:sigmaztime} for the run with isotropic CR transport and $\kappa = 3 \times 10^{27}$~cm$^2$~s$^{-1}$. The two upper panels show the vertical scale height for the different phases as in figure \ref{Fig:phasescaleheight}, but for this one run at four different moments in its recent history (at 210, 227, 246, and 251~Myr of age). These moments are circled in the SFR curve in the third panel. The bottom panel shows the radial profiles of gas mass surface density and the S\'ersic fits to these profiles.

 The large-scale gas distributions are found to be very stable over the last 50 Myr. The same conclusion is reached for the other simulations. 
 
\end{appendix} 

\end{document}